\documentstyle[epsfig]{mn2e}
\epsfverbosetrue

\title[Chemical abundances for Hf\,2-2]
{Chemical abundances for Hf~2-2, a planetary nebula with the strongest 
known heavy element recombination lines}
\author[X.-W. Liu et al.]
{X.-W. Liu$^{1}$, M. J. Barlow$^2$, Y. Zhang$^1$, R. J. Bastin$^2$, P. J. Storey$^2$%, I.J. Danziger$^3$
\\
$^1$ Department of Astronomy, Peking University, Beijing 100871, P.R. China\\
$^2$ Department of Physics and Astronomy, University College London,
      Gower Street, London WC1E 6BT, UK\\
}
\date{Received:}

\begin{document}

\maketitle

\begin{abstract}

We present high quality optical spectroscopic observations of the planetary
nebula (PN) Hf\,2-2. The spectrum exhibits many prominent optical recombination
lines (ORLs) from heavy element ions. Analysis of the H~{\sc i} and He~{\sc i}
recombination spectrum yields an electron temperature of $\sim 900$~K, a factor
of ten lower than given by the collisionally excited [O~{\sc iii}] forbidden
lines. The ionic abundances of heavy elements relative to hydrogen derived from
ORLs are about a factor of 70 higher than those deduced from collisionally
excited lines (CELs) from the same ions, the largest abundance discrepancy
factor (adf) ever measured for a PN. By comparing the observed O~{\sc ii}
$\lambda$4089/$\lambda$4649 ORL ratio to theoretical value as a function of
electron temperature,
% calculated down to a temperature of 300~K, 
we show that the O~{\sc ii} ORLs arise from ionized regions with an electron
temperature of only $\sim 630$~K. The current observations thus provide the
strongest evidence that the nebula contains another previously unknown
component of cold, high metallicity gas, which is too cool to excite any
significant optical or UV CELs and is thus invisible via such lines.  The
existence of such a plasma component in PNe provides a natural solution to the
long-standing dichotomy between nebular plasma diagnostics and abundance
determinations using CELs on the one hand and ORLs on the other.

\end{abstract}

%\nokeywords

\section{Introduction} 

In recent years a number of papers have been published reporting very deep
spectroscopic observations of the relatively faint optical recombination lines
(ORLs) emitted by heavy element ions in planetary nebulae (PNe), e.g. Liu et
al. (1995, 2000, 2001, 2004a,b), Garnett \& Dinerstein (2001), Ruiz et al.
(2003), Tsamis et al. (2003b, 2004), Peimbert et al. (2004), Sharpee, Baldwin
\& Williams (2004), Wesson, Liu \& Barlow (2005). A common feature of the
analyses of these ORLs was that they yielded systematically higher ionic
abundances than obtained for the same ions from classical nebular forbidden
lines (also known as collisionally excited lines, or CELs). For most PNe, the
ORL/CEL abundance discrepancy factors (adf's) typically lie in the range
1.6--3.0, but with a significant tail extending to much higher adf values. Liu
et al. (1995) and Luo, Liu \& Barlow (2001) found a mean (with respect to all
the heavy elements measured) adf of $\sim 5$ for NGC 7009, while Liu et al.
(2000) derived a mean adf of $\sim 10$ for NGC 6153 and Liu et al. (2001)
obtained mean adf's of $\sim 6$ and 20 for the bulge PNe M~2-36 and
M~1-42, respectively. Kaler (1988) reported an extremely strong C~{\sc ii}
$\lambda$4267 recombination line from the southern PN Hf~2-2, with an observed
intensity of 9--10 on a scale where H$\beta$ = 100, which represents an
enhancement of more than a factor of ten compared to most PNe. As part of a
search for further high-adf nebulae, we therefore observed Hf~2-2 and report
our results here, which confirm that this nebula has the strongest ORLs and
highest adf's known for any planetary nebula.

\section{Observations} 

\begin{figure*} \centering \epsfig{file=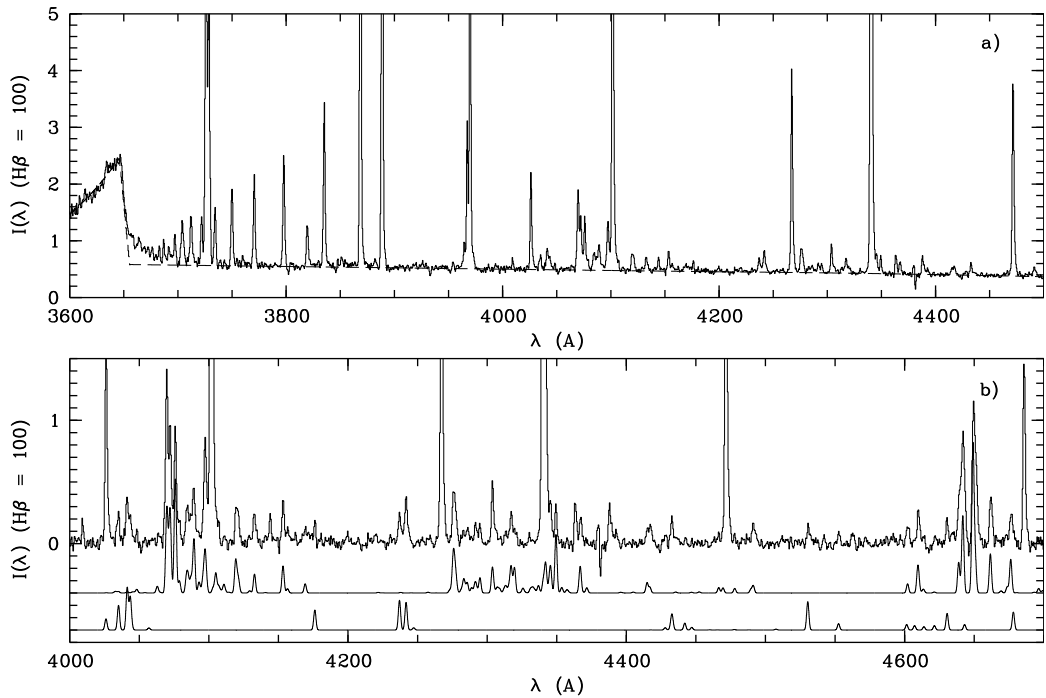, height=14.0cm, bbllx=64pt,
bblly=30pt, bburx=550pt, bbury=768pt, clip=, angle=270}  \caption {Spectrum of
Hf\,2-2 obtained with the 2\,arcsec wide slit. The spectrum has been corrected
for interstellar reddening and normalized such that H$\beta$ has an integrated
flux of 100. a) $\lambda\lambda$3600--4500 showing the huge Balmer jump at
3645\,{\AA} that declines steeply towards shorter wavelengths, indicating an
average electron temperature of 900~K for the hydrogen recombination spectrum.
Overplotted is an empirical fit of the continuum (stellar plus nebular); b)
$\lambda\lambda$4000--4700 showing the swarm of, often blended, recombination
lines from C, N, O and Ne ions. The continuum has been subtracted. Also
overplotted are synthesized recombination line spectra of O~{\sc ii} (middle
curve) and N~{\sc ii} (bottom curve), shifted downward by 0.4 and 0.7,
respectively. They were calculated assuming $T_{\rm e} = 900$\,K, $\log N_{\rm
e} = 3$ (cm$^{-3}$), $N({\rm O}^{2+})/N({\rm H}^+) = 7.48\times 10^{-3}$ and
$N({\rm N}^{2+})/N({\rm H}^+) = 2.72\times 10^{-3}$ (c.f. Table~4).}
\end{figure*}

Hf\,2-2 was observed on three photometric nights in June 2001 with the
Boller \& Chivens long-slit spectrograph mounted on the ESO 1.52-m
telescope. Two wavelength regions were observed. The seeing was between
2--3\,arcsec, throughout most of the night of June 18, followed by three
nights of sub-arcsec seeing. The slit was positioned at PA$=45^{\rm o}$,
centred on the central star. The CCD was a 2672$\times$512 Loral
UV-flooded chip with a pixel size of 15~$\mu$m and a read-out-noise of
7.2\,e$^-$\,pix$^{-1}$ rms.

Two wavelength regions were observed, see Table~1 for a journal of
observations. The blue region was observed with a 2400\,lines\,mm$^{-1}$
holographic grating, which yielded a dispersion of 32\,{\AA}\,mm$^{-1}$, or
$\sim 0.5$\,{\AA}\,pix$^{-1}$. Some vignetting was observed, in particular on
the blue side of this large format CCD, which limited the useful wavelength
range to 3540--4800~{\AA}. A slitwidth of 2\,arcsec was used for maximum
spectral resolution. In addition, in order to obtain magnitudes for the central
star, two spectra with slitwidths of 4 and 8\,arcsec were also obtained.
Because of the UV flooding procedure to maximise sensitivity, the instrumental
line width was significantly broader than predicted by the instrumental optics
alone -- a 2\,arcsec slit projects to only 1.32\,pix on the CCD, yet the actual
measured line width was about 3\,pix, or 1.5\,{\AA} FWHM.  The spectral
resolution degraded even more towards the edges of the spectral coverage,
approaching 2\,{\AA} FWHM. The red wavelength region, from 4750--7230\,{\AA},
was observed with a 1200\,lines\,mm$^{-1}$ ruled grating at a dispersion of
65.6\,{\AA}\,mm$^{-1}$. For a 2\,arcsec wide slit, this yielded a spectral
resolution of 2.8\,{\AA} FWHM including the effects caused by the UV-flooding.
All spectra were wavelength-calibrated using exposures of a He-Ar-Ne-Fe lamp
and flux-calibrated using 8\,arcsec wide slit observations of the {\it HST}
standard stars Feige 110 and the central star of the PN NGC\,7293. Example
spectra obtained with a 2\,arcsec slitwidth are shown in Fig.\,1.

\begin{table}
\begin{minipage}{70mm}
\centering
\caption{Journal of observations}
\begin{tabular}{ccccc}
Date & $\lambda$-range & FWHM & Slit width &Exp. Time\\
(UT) & ({\AA}) & ({\AA}) & (arcsec) & (sec)\\
\noalign{\vskip3pt}
18/06/01& 3550-4800 & 1.5 &  2 & $4\times 2400$\\
18/06/01& 3550-4800 & 2.6 &  8 & 1800\\
19/06/01& 3550-4800 & 1.5 &  2 & 1800\\
19/06/01& 3550-4800 & 1.9 &  4 & 1800\\
21/06/01& 4750-7230 & 2.8 &  2 & $4\times 1800$\\
21/06/01& 4750-7230 & 5.8 &  8 & 1800\\
05/08/01& 4000-7700 & 8.0 &  4 &  600\\
\end{tabular}
\end{minipage}
\end{table}

\begin{figure*} \centering \epsfig{file=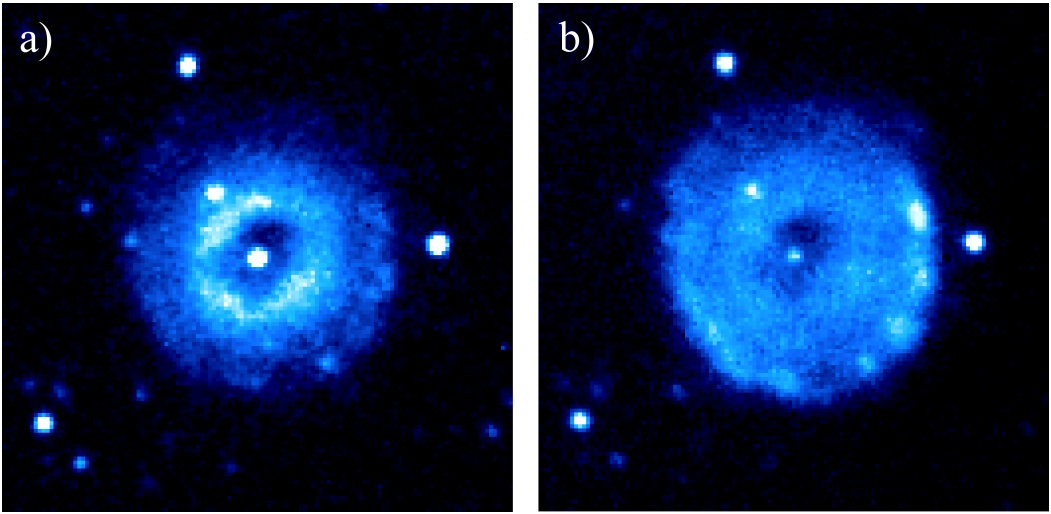, width=14.0cm, bbllx=32pt,
bblly=40pt, bburx=548pt, bbury=292pt, clip=, angle=0}  \caption {Narrow
band images of Hf\,2-2 in the light of a) [O~{\sc iii}] $\lambda$5007
and b) H$\alpha$, obtained by Schwarz et al. (1992) with the 3.5-m
NTT. The field of view is 33.5$\times$33.5\,arcsec$^2$. 
North is up and east to the left.}
\end{figure*}

On the night of August 5 2001, an additional spectrum of Hf\,2-2 was
obtained with the IDS Spectrograph mounted on the 2.5-m Isaac Newton
Telescope (INT) at La Palma Observatory. The spectrum was obtained with a
316\,lines\,mm$^{-1}$ ruled grating in first spectral order, covering the
4000--7700~{\AA} wavelength range at a resolution of 8~{\AA} FWHM. A GG385
order sorting filter was used to block out second order light. The CCD was
a $4096\times512$ Loral chip with 13.5~$\mu$m pixels. A slitwidth of
4\,arcsec was used. The spectrum was calibrated using observations of the
{\it HST} standard star BD\,$+$28$^{\rm o}$\,4211. The INT spectrum was
particularly useful for joining the blue and red spectra obtained at ESO
without relying on their absolute flux calibration. In addition, the flux
of H$\beta$ in the ESO red spectra was unreliable as it fell close the
edge of the spectral coverage and was partially affected by vignetting.

\section{Absolute H$\beta$ flux} 

Hf\,2-2 was imaged in the light of [O~{\sc iii}] $\lambda$5007 and H$\alpha$ by
Schwarz, Corradi \& Melnick (1992) during their imaging survey of southern
hemisphere PNe. The images were obtained with the BFOCS2 instrument mounted on
the 3.5-m New Technology Telescope (NTT), with a spatial sampling of
0.2594\,arcsec\,pix$^{-1}$. These images are reproduced in Fig.~2. They have
been rotated to the normal orientation such that North is up and East to the
left. Both the [O~{\sc iii}] and H$\alpha$ images show that Hf\,2-2 is highly
symmetric, with an inner disk-like shell and an outer limb-brightened shell
that has an angular diameter of 25\,arcs. The inner disk (shell) is brighter in
[O~{\sc iii}] and has a central cavity, which is partially filled at
PA$=60^{\rm o}$ leaving two holes lying symmetrically about the central star.

The ESO red spectra obtained with the 2 and 8\,arcsec slitwidths yielded
H$\alpha$ fluxes of 5.42 and $23.0\times 10^{-13}$\,ergs\,cm$^{-2}$\,s$^{-1}$,
respectively, after integration along the slit. Since the H$\beta$ fluxes from
the ESO red spectra were unreliable due to vignetting, we converted the
H$\alpha$ fluxes to those for H$\beta$ using the observed ratio of $F({\rm
H}\alpha)/F({\rm H}\beta) = 4.21$ derived from the INT spectrum, which implies
$F({\rm H}\beta)$ = 1.29 and $5.47\times 10^{-13}$\,ergs\,cm$^{-2}$\,s$^{-1}$
for the ESO 2 and 8\,arcsec slitwidth observations, respectively. To obtain the
H$\beta$ flux for the whole nebula. we made use of the H$\alpha$ image of
Schwarz et al. We found that our 2 and 8\,arcsec slits should have respectively
caught 12.5 per cent and 48.8 per cent of the H$\beta$ flux emitted by Hf\,2-2,
implying total H$\beta$ fluxes of 1.03 and $1.12\times
10^{-12}$\,ergs\,cm$^{-2}$\,s$^{-1}$, or ${\rm log}\,F({\rm H}\beta) = -11.99$
and $-11.95$ (ergs\,cm$^{-2}$\,s$^{-1}$), respectively. We shall adopt ${\rm
log}\,F({\rm H}\beta) = -11.95$ (ergs\,cm$^{-2}$\,s$^{-1}$) for our analysis, a
value which is probably accurate to 0.05\,dex.

\section{Interstellar extinction} 

The amount of interstellar reddening towards Hf\,2-2 has to be determined
before the observed line fluxes can be used to study the thermal and
density structure of the nebula and its elemental abundances. However,
apart from its very prominent heavy element ORL spectra, the most striking
feature observed in the optical spectrum of Hf\,2-2 is its very large
Balmer continuum jump, accompanied by a steep decline in the continuum
intensity on the blue wavelength side of the discontinuity, indicating a
{\it very} low electron temperature -- around 1000~K from a crude
estimate, about an order of magnitude lower than the normal value of 10\
000~K found in a typical photoionized gaseous nebula. The temperature
implied by the Balmer jump is so low that the small but significant
dependence of the Balmer line decrement on electron temperature must be
taken into account before it can be used to derive the extinction towards
Hf\,2-2.

\setcounter{table}{1}
\begin{table*}
\begin{minipage}{160mm}
\centering
\caption{Observed relative line fluxes, on a scale where H$\beta = 100$}
\begin{tabular}{lccrrrrrr}
{Ident     } & $\lambda$ & $f(\lambda)$ & \multicolumn{3}{c}{Observed fluxes} &  \multicolumn{3}{c}{Dereddened fluxes} \\
{          } & ({\AA})   &              & 2$\arcsec$ slitwidth & 4$\arcsec$ slitwidth &  8$\arcsec$ slitwidth &  2$\arcsec$ slitwidth &  4$\arcsec$ slitwidth &  8$\arcsec$ slitwidth \\
{Continuum } & 3643 &  0.270 &   2.20          &   2.04          &   1.86          &   2.49          &   2.31          &   2.11          \\
{Continuum } & 3678 &  0.265 &   0.54          &   0.45          &   0.28          &   0.61          &   0.50          &   0.31          \\
{H 20      } & 3683 &  0.264 &   0.73$\pm$0.10 &   0.74$\pm$0.17 &   0.62$\pm$0.10 &   0.82$\pm$0.12 &   0.84$\pm$0.19 &   0.71$\pm$0.11 \\
{H 19      } & 3687 &  0.263 &   0.92$\pm$0.11 &   0.81$\pm$0.17 &   0.79$\pm$0.10 &   1.04$\pm$0.13 &   0.92$\pm$0.20 &   0.89$\pm$0.12 \\
{H 18      } & 3692 &  0.262 &   0.80$\pm$0.11 &   0.82$\pm$0.17 &   0.87$\pm$0.11 &   0.90$\pm$0.12 &   0.92$\pm$0.19 &   0.98$\pm$0.12 \\
{H 17      } & 3697 &  0.262 &   1.18$\pm$0.12 &   1.25$\pm$0.20 &   0.93$\pm$0.11 &   1.33$\pm$0.14 &   1.41$\pm$0.23 &   1.05$\pm$0.12 \\
{H16,HeI   } & 3704 &  0.260 &   2.18$\pm$0.18 &   2.10$\pm$0.30 &   1.90$\pm$0.17 &   2.45$\pm$0.21 &   2.37$\pm$0.33 &   2.14$\pm$0.19 \\
{H 15      } & 3712 &  0.259 &   2.16$\pm$0.17 &   2.14$\pm$0.26 &   1.88$\pm$0.17 &   2.43$\pm$0.19 &   2.42$\pm$0.30 &   2.12$\pm$0.19 \\
{H14,[SIII]} & 3722 &  0.257 &   1.65$\pm$0.18 &   1.98$\pm$0.32 &   1.35$\pm$0.24 &   1.86$\pm$0.20 &   2.22$\pm$0.36 &   1.52$\pm$0.27 \\
{[O II]    } & 3726 &  0.257 &  14.48$\pm$0.69 &  14.58$\pm$0.42 &  13.84$\pm$1.06 &  16.29$\pm$0.77 &  16.41$\pm$0.48 &  15.58$\pm$1.19 \\
{[O II]    } & 3729 &  0.256 &   9.75$\pm$0.57 &  10.99$\pm$0.38 &   9.38$\pm$1.06 &  10.97$\pm$0.64 &  12.37$\pm$0.43 &  10.56$\pm$1.19 \\
{H 13      } & 3734 &  0.255 &   2.04$\pm$0.13 &   1.64$\pm$0.18 &   1.74$\pm$0.14 &   2.29$\pm$0.14 &   1.84$\pm$0.20 &   1.96$\pm$0.15 \\
{H 12      } & 3750 &  0.253 &   2.72$\pm$0.13 &   3.13$\pm$0.18 &   2.83$\pm$0.13 &   3.05$\pm$0.14 &   3.51$\pm$0.20 &   3.18$\pm$0.14 \\
{O III     } & 3760 &  0.251 &   0.35$\pm$0.14 &                 &                 &   0.39$\pm$0.15 &                 &                 \\
{H 11      } & 3770 &  0.249 &   3.03$\pm$0.13 &   2.80$\pm$0.18 &   2.85$\pm$0.13 &   3.40$\pm$0.14 &   3.14$\pm$0.20 &   3.20$\pm$0.14 \\
{H 10      } & 3797 &  0.244 &   3.84$\pm$0.13 &   4.49$\pm$0.18 &   3.58$\pm$0.13 &   4.29$\pm$0.14 &   5.03$\pm$0.20 &   4.01$\pm$0.14 \\
{He I      } & 3819 &  0.240 &   1.53$\pm$0.14 &   1.72$\pm$0.18 &   1.58$\pm$0.13 &   1.71$\pm$0.15 &   1.92$\pm$0.20 &   1.77$\pm$0.14 \\
{H 9       } & 3835 &  0.237 &   5.45$\pm$0.13 &   5.46$\pm$0.17 &   5.42$\pm$0.13 &   6.08$\pm$0.14 &   6.09$\pm$0.19 &   6.05$\pm$0.14 \\
{[Ne III]  } & 3867 &  0.231 &  20.29$\pm$0.25 &  20.18$\pm$0.25 &  19.23$\pm$0.25 &  22.56$\pm$0.28 &  22.45$\pm$0.28 &  21.39$\pm$0.28 \\
{H8,HeI    } & 3888 &  0.227 &  18.81$\pm$0.20 &  17.96$\pm$0.30 &  18.49$\pm$0.27 &  20.88$\pm$0.22 &  19.94$\pm$0.33 &  20.53$\pm$0.30 \\
{He I      } & 3964 &  0.211 &   0.83$\pm$0.09 &   0.78$\pm$0.10 &   0.65$\pm$0.16 &   0.92$\pm$0.10 &   0.86$\pm$0.11 &   0.72$\pm$0.18 \\
{[Ne III]  } & 3967 &  0.211 &   4.37$\pm$0.11 &   4.36$\pm$0.12 &   4.14$\pm$0.19 &   4.82$\pm$0.12 &   4.81$\pm$0.13 &   4.56$\pm$0.21 \\
{H 7       } & 3970 &  0.210 &  12.15$\pm$0.16 &  12.47$\pm$0.18 &  12.68$\pm$0.26 &  13.39$\pm$0.17 &  13.73$\pm$0.20 &  13.97$\pm$0.29 \\
{He I      } & 4009 &  0.202 &   0.31$\pm$0.06 &   0.31$\pm$0.08 &                 &   0.34$\pm$0.07 &   0.35$\pm$0.09 &                 \\
{He I      } & 4026 &  0.198 &   3.17$\pm$0.11 &   0.31$\pm$0.15 &   3.20$\pm$0.11 &   3.47$\pm$0.12 &   0.34$\pm$0.16 &   3.51$\pm$0.12 \\
{N II      } & 4035 &  0.196 &   0.48$\pm$0.06 &   0.37$\pm$0.08 &   0.31$\pm$0.06 &   0.53$\pm$0.07 &   0.40$\pm$0.09 &   0.34$\pm$0.06 \\
{N II      } & 4041 &  0.195 &   0.72$\pm$0.07 &   0.63$\pm$0.09 &   0.56$\pm$0.08 &   0.79$\pm$0.07 &   0.69$\pm$0.10 &   0.61$\pm$0.09 \\
{N II      } & 4044 &  0.194 &   0.45$\pm$0.06 &   0.33$\pm$0.09 &   0.46$\pm$0.08 &   0.49$\pm$0.07 &   0.36$\pm$0.10 &   0.50$\pm$0.09 \\
{O II 69.62} & 4070 &  0.189 &   1.06$\pm$0.03 &   1.03$\pm$0.05 &   0.97$\pm$0.04 &   1.15$\pm$0.03 &   1.13$\pm$0.05 &   1.05$\pm$0.04 \\
{O II 69.89} & 4070 &  0.189 &   1.69$\pm$0.05 &   1.65$\pm$0.07 &   1.54$\pm$0.05 &   1.84$\pm$0.06 &   1.80$\pm$0.08 &   1.68$\pm$0.06 \\
{O II 72.16} & 4072 &  0.188 &   1.88$\pm$0.10 &   1.82$\pm$0.14 &   1.58$\pm$0.12 &   2.05$\pm$0.10 &   1.98$\pm$0.15 &   1.73$\pm$0.13 \\
{O II 75.86} & 4076 &  0.187 &   1.85$\pm$0.08 &   1.65$\pm$0.12 &   1.69$\pm$0.08 &   2.02$\pm$0.09 &   1.80$\pm$0.13 &   1.84$\pm$0.09 \\
{O II 78.84} & 4079 &  0.187 &   0.41$\pm$0.07 &   0.32$\pm$0.10 &   0.36$\pm$0.07 &   0.45$\pm$0.08 &   0.35$\pm$0.10 &   0.39$\pm$0.07 \\
{O II 83.90} & 4084 &  0.186 &   0.41$\pm$0.09 &   0.30$\pm$0.12 &   0.25$\pm$0.10 &   0.45$\pm$0.09 &   0.32$\pm$0.14 &   0.27$\pm$0.11 \\
{O II 85.11} & 4085 &  0.185 &   0.39$\pm$0.09 &   0.56$\pm$0.13 &   0.51$\pm$0.11 &   0.42$\pm$0.10 &   0.61$\pm$0.14 &   0.56$\pm$0.12 \\
{O II 87.15} & 4087 &  0.185 &   0.42$\pm$0.07 &   0.21$\pm$0.10 &   0.20$\pm$0.09 &   0.46$\pm$0.08 &   0.23$\pm$0.11 &   0.22$\pm$0.09 \\
{O II 89.29} & 4089 &  0.184 &   0.93$\pm$0.07 &   0.90$\pm$0.10 &   0.92$\pm$0.08 &   1.01$\pm$0.08 &   0.98$\pm$0.11 &   1.01$\pm$0.08 \\
{O II blend} & 4097 &  0.183 &   2.24$\pm$0.20 &   2.03$\pm$0.18 &   1.78$\pm$0.35 &   2.44$\pm$0.22 &   2.21$\pm$0.20 &   1.93$\pm$0.38 \\
{H I       } & 4101 &  0.182 &  21.03$\pm$0.55 &  21.66$\pm$0.51 &  21.77$\pm$0.35 &  22.86$\pm$0.60 &  23.55$\pm$0.55 &  23.67$\pm$0.38 \\
{O II blend} & 4120 &  0.178 &   0.90$\pm$0.10 &   1.05$\pm$0.14 &   0.91$\pm$0.10 &   0.98$\pm$0.11 &   1.14$\pm$0.15 &   0.99$\pm$0.10 \\
{O II      } & 4133 &  0.175 &   0.56$\pm$0.06 &   0.31$\pm$0.07 &   0.41$\pm$0.06 &   0.61$\pm$0.07 &   0.33$\pm$0.08 &   0.45$\pm$0.06 \\
{He I      } & 4144 &  0.172 &   0.47$\pm$0.06 &   0.42$\pm$0.08 &   0.54$\pm$0.06 &   0.51$\pm$0.06 &   0.45$\pm$0.08 &   0.58$\pm$0.07 \\
{O II      } & 4153 &  0.170 &   0.74$\pm$0.07 &   0.83$\pm$0.10 &   0.84$\pm$0.08 &   0.80$\pm$0.08 &   0.89$\pm$0.11 &   0.91$\pm$0.08 \\
{O II      } & 4157 &  0.169 &   0.27$\pm$0.05 &   0.40$\pm$0.08 &   0.14$\pm$0.06 &   0.30$\pm$0.06 &   0.43$\pm$0.08 &   0.15$\pm$0.06 \\
{O II      } & 4169 &  0.167 &   0.31$\pm$0.05 &   0.19$\pm$0.07 &   0.36$\pm$0.05 &   0.33$\pm$0.06 &   0.21$\pm$0.07 &   0.39$\pm$0.06 \\
{N II      } & 4176 &  0.165 &   0.36$\pm$0.05 &   0.32$\pm$0.07 &   0.35$\pm$0.05 &   0.39$\pm$0.06 &   0.35$\pm$0.08 &   0.38$\pm$0.06 \\
{NeII blend} & 4220 &  0.155 &   0.22$\pm$0.04 &   0.27$\pm$0.07 &                 &   0.23$\pm$0.05 &   0.29$\pm$0.08 &                 \\
{N II blend} & 4237 &  0.151 &   0.61$\pm$0.08 &   0.65$\pm$0.13 &   0.56$\pm$0.08 &   0.65$\pm$0.08 &   0.69$\pm$0.14 &   0.60$\pm$0.09 \\
{N II blend} & 4242 &  0.150 &   1.00$\pm$0.08 &   0.80$\pm$0.12 &   0.86$\pm$0.09 &   1.07$\pm$0.08 &   0.86$\pm$0.13 &   0.92$\pm$0.09 \\
{C II      } & 4267 &  0.144 &   7.11$\pm$0.19 &   7.09$\pm$0.19 &   6.71$\pm$0.15 &   7.60$\pm$0.20 &   7.58$\pm$0.20 &   7.17$\pm$0.16 \\
{O II blend} & 4276 &  0.142 &   1.37$\pm$0.08 &   1.44$\pm$0.15 &   1.22$\pm$0.10 &   1.47$\pm$0.09 &   1.53$\pm$0.16 &   1.30$\pm$0.10 \\
{O II blend} & 4383 &  0.117 &   0.23$\pm$0.04 &   0.30$\pm$0.06 &   0.67$\pm$0.10 &   0.24$\pm$0.05 &   0.31$\pm$0.07 &   0.71$\pm$0.10 \\
{O II      } & 4286 &  0.140 &   0.29$\pm$0.04 &   0.32$\pm$0.06 &   $\uparrow$    &   0.31$\pm$0.05 &   0.34$\pm$0.07 &   $\uparrow$    \\
{O II blend} & 4292 &  0.138 &   0.36$\pm$0.05 &   0.24$\pm$0.06 &   0.86$\pm$0.08 &   0.38$\pm$0.05 &   0.25$\pm$0.07 &   0.91$\pm$0.09 \\
{O II blend} & 4295 &  0.137 &   0.32$\pm$0.04 &   0.36$\pm$0.06 &   $\uparrow$    &   0.34$\pm$0.05 &   0.38$\pm$0.07 &   $\uparrow$    \\
{O II blend} & 4304 &  0.135 &   1.00$\pm$0.06 &   1.02$\pm$0.10 &   1.24$\pm$0.09 &   1.07$\pm$0.07 &   1.09$\pm$0.10 &   1.32$\pm$0.10 \\
{O II + ?  } & 4306 &  0.135 &   0.23$\pm$0.05 &   0.24$\pm$0.07 &   $\uparrow$    &   0.25$\pm$0.05 &   0.26$\pm$0.07 &   $\uparrow$    \\
{O II blend} & 4317 &  0.132 &   0.56$\pm$0.05 &   0.57$\pm$0.08 &   0.78$\pm$0.11 &   0.59$\pm$0.05 &   0.60$\pm$0.08 &   0.83$\pm$0.11 \\
{O II      } & 4320 &  0.132 &   0.26$\pm$0.04 &   0.27$\pm$0.07 &   $\uparrow$    &   0.27$\pm$0.05 &   0.29$\pm$0.07 &   $\uparrow$    \\
{H I       } & 4340 &  0.127 &  42.90$\pm$1.16 &  43.22$\pm$0.95 &  42.79$\pm$0.53 &  45.48$\pm$1.23 &  45.82$\pm$1.01 &  45.37$\pm$0.56 \\
{O II      } & 4346 &  0.125 &   0.75$\pm$0.08 &   0.67$\pm$0.08 &   0.58$\pm$0.07 &   0.79$\pm$0.09 &   0.71$\pm$0.09 &   0.62$\pm$0.07 \\
{O II      } & 4349 &  0.125 &   0.62$\pm$0.08 &   0.87$\pm$0.09 &   0.66$\pm$0.06 &   0.65$\pm$0.08 &   0.92$\pm$0.09 &   0.70$\pm$0.06 \\
{[O III]   } & 4363 &  0.121 &   0.70$\pm$0.08 &   0.83$\pm$0.09 &   0.77$\pm$0.06 &   0.74$\pm$0.08 &   0.88$\pm$0.09 &   0.82$\pm$0.07 \\
{O II      } & 4367 &  0.120 &   0.46$\pm$0.07 &   0.69$\pm$0.08 &   0.53$\pm$0.06 &   0.48$\pm$0.08 &   0.73$\pm$0.09 &   0.57$\pm$0.06 \\
\end{tabular}
\end{minipage}
\end{table*}

\setcounter{table}{1}
\begin{table*}
\begin{minipage}{160mm}
\centering
\caption{{\it -- continued}}
\begin{tabular}{lccrrrrrr}
{Ident     } & $\lambda$ & $f(\lambda)$ & \multicolumn{3}{c}{Observed fluxes} &  \multicolumn{3}{c}{Dereddened fluxes} \\
{          } & ({\AA})   &              & 2$\arcsec$ slitwidth & 4$\arcsec$ slitwidth &  8$\arcsec$ slitwidth &  2$\arcsec$ slitwidth &  4$\arcsec$ slitwidth &  8$\arcsec$ slitwidth \\
{N III     } & 4379 &  0.118 &   0.23$\pm$0.07 &   0.31$\pm$0.07 &   0.39$\pm$0.05 &   0.25$\pm$0.07 &   0.33$\pm$0.08 &   0.41$\pm$0.06 \\
{He I      } & 4387 &  0.116 &   0.67$\pm$0.08 &   0.75$\pm$0.08 &   0.91$\pm$0.07 &   0.71$\pm$0.08 &   0.79$\pm$0.09 &   0.96$\pm$0.07 \\
{Ne II     } & 4392 &  0.115 &   0.21$\pm$0.07 &   0.22$\pm$0.07 &   0.29$\pm$0.05 &   0.22$\pm$0.07 &   0.23$\pm$0.07 &   0.31$\pm$0.06 \\
{O II      } & 4415 &  0.109 &   0.26$\pm$0.08 &   0.21$\pm$0.07 &   0.26$\pm$0.09 &   0.27$\pm$0.08 &   0.22$\pm$0.07 &   0.27$\pm$0.09 \\
{O II      } & 4417 &  0.109 &   0.32$\pm$0.08 &   0.23$\pm$0.07 &   0.27$\pm$0.08 &   0.33$\pm$0.08 &   0.24$\pm$0.07 &   0.28$\pm$0.09 \\
{N II      } & 4433 &  0.105 &   0.48$\pm$0.07 &   0.42$\pm$0.05 &   0.48$\pm$0.05 &   0.51$\pm$0.07 &   0.44$\pm$0.05 &   0.51$\pm$0.05 \\
{He I      } & 4471 &  0.096 &   7.54$\pm$0.18 &   7.52$\pm$0.15 &   7.43$\pm$0.10 &   7.88$\pm$0.19 &   7.86$\pm$0.15 &   7.76$\pm$0.10 \\
{O II, C II} & 4491 &  0.091 &   0.47$\pm$0.06 &   0.45$\pm$0.08 &   0.59$\pm$0.08 &   0.49$\pm$0.06 &   0.47$\pm$0.08 &   0.61$\pm$0.08 \\
{N II      } & 4530 &  0.081 &   0.38$\pm$0.05 &   0.42$\pm$0.08 &   0.37$\pm$0.06 &   0.39$\pm$0.06 &   0.43$\pm$0.08 &   0.38$\pm$0.06 \\
{N II      } & 4552 &  0.076 &   0.22$\pm$0.05 &   0.27$\pm$0.06 &   0.24$\pm$0.05 &   0.23$\pm$0.05 &   0.28$\pm$0.07 &   0.25$\pm$0.06 \\
{Mg I]     } & 4562 &  0.073 &   0.22$\pm$0.05 &   0.27$\pm$0.06 &   0.25$\pm$0.05 &   0.23$\pm$0.05 &   0.27$\pm$0.07 &   0.26$\pm$0.06 \\
{O II      } & 4602 &  0.064 &   0.40$\pm$0.09 &   0.34$\pm$0.09 &   0.33$\pm$0.08 &   0.41$\pm$0.10 &   0.35$\pm$0.09 &   0.34$\pm$0.08 \\
{O II      } & 4609 &  0.062 &   0.66$\pm$0.09 &   0.79$\pm$0.10 &   0.65$\pm$0.09 &   0.68$\pm$0.09 &   0.81$\pm$0.10 &   0.67$\pm$0.10 \\
{N II      } & 4630 &  0.057 &   0.45$\pm$0.06 &   0.57$\pm$0.09 &   0.40$\pm$0.06 &   0.46$\pm$0.07 &   0.59$\pm$0.10 &   0.41$\pm$0.06 \\
{N III     } & 4634 &  0.056 &   0.26$\pm$0.06 &   0.32$\pm$0.09 &   0.28$\pm$0.06 &   0.26$\pm$0.07 &   0.33$\pm$0.09 &   0.29$\pm$0.06 \\
{O II      } & 4639 &  0.055 &   1.03$\pm$0.09 &   1.01$\pm$0.12 &   0.84$\pm$0.09 &   1.05$\pm$0.09 &   1.03$\pm$0.12 &   0.86$\pm$0.10 \\
{N III     } & 4641 &  0.054 &   0.40$\pm$0.14 &   0.14$\pm$0.20 &   0.73$\pm$0.16 &   0.41$\pm$0.14 &   0.14$\pm$0.20 &   0.75$\pm$0.17 \\
{O II      } & 4642 &  0.054 &   2.14$\pm$0.35 &   2.38$\pm$0.50 &   1.63$\pm$0.33 &   2.20$\pm$0.36 &   2.44$\pm$0.51 &   1.67$\pm$0.34 \\
{N III     } & 4642 &  0.054 &   0.05$\pm$0.01 &   0.06$\pm$0.02 &   0.06$\pm$0.01 &   0.05$\pm$0.01 &   0.07$\pm$0.02 &   0.06$\pm$0.01 \\
{O II      } & 4649 &  0.052 &   2.68$\pm$0.13 &   2.62$\pm$0.17 &   2.68$\pm$0.14 &   2.75$\pm$0.13 &   2.68$\pm$0.17 &   2.75$\pm$0.14 \\
{O II      } & 4651 &  0.052 &   1.23$\pm$0.13 &   1.10$\pm$0.16 &   0.86$\pm$0.16 &   1.26$\pm$0.13 &   1.13$\pm$0.16 &   0.88$\pm$0.16 \\
{O II      } & 4662 &  0.049 &   1.04$\pm$0.07 &   1.19$\pm$0.11 &   0.94$\pm$0.06 &   1.06$\pm$0.07 &   1.22$\pm$0.11 &   0.96$\pm$0.06 \\
{O II      } & 4674 &  0.046 &   0.08$\pm$0.00 &   0.07$\pm$0.00 &   0.08$\pm$0.00 &   0.08$\pm$0.00 &   0.07$\pm$0.00 &   0.08$\pm$0.00 \\
{O II      } & 4675 &  0.046 &   0.42$\pm$0.02 &   0.39$\pm$0.02 &   0.43$\pm$0.02 &   0.43$\pm$0.02 &   0.40$\pm$0.02 &   0.44$\pm$0.02 \\
{N II      } & 4678 &  0.045 &   0.31$\pm$0.06 &   0.16$\pm$0.09 &   0.29$\pm$0.06 &   0.32$\pm$0.07 &   0.17$\pm$0.09 &   0.29$\pm$0.06 \\
{He II     } & 4686 &  0.043 &   3.13$\pm$0.10 &   3.06$\pm$0.17 &   2.73$\pm$0.11 &   3.19$\pm$0.10 &   3.13$\pm$0.17 &   2.78$\pm$0.11 \\
{He I      } & 4713 &  0.036 &   0.48$\pm$0.05 &   0.42$\pm$0.10 &   0.52$\pm$0.06 &   0.49$\pm$0.05 &   0.43$\pm$0.10 &   0.53$\pm$0.06 \\
{H I       } & 4861 &  0.000 & 100.00          &                 & 100.00          & 100.00          &                 & 100.00          \\
{He I      } & 4921 & -0.015 &   2.83$\pm$0.26 &                 &   2.66$\pm$0.20 &   2.81$\pm$0.26 &                 &   2.64$\pm$0.20 \\
{[O III]   } & 4959 & -0.024 &  61.70$\pm$0.40 &  63.90$\pm$3.62 &  62.40$\pm$2.00 &  61.02$\pm$0.40 &  63.20$\pm$3.58 &  61.71$\pm$1.98 \\
{[O III]   } & 5007 & -0.036 & 191.00$\pm$2.00 & 198.00$\pm$3.63 & 197.00$\pm$4.47 & 187.88$\pm$1.97 & 194.74$\pm$3.57 & 193.78$\pm$4.40 \\
{He I      } & 5015 & -0.038 &   3.08$\pm$0.14 &                 &   $\uparrow$    &   3.03$\pm$0.14 &                 &   $\uparrow$    \\
{[N I]     } & 5200 & -0.083 &   1.37$\pm$0.10 &                 &   1.42$\pm$0.10 &   1.32$\pm$0.10 &                 &   1.37$\pm$0.10 \\
{C II      } & 5342 & -0.117 &   0.33$\pm$0.09 &                 &   0.29$\pm$0.09 &   0.31$\pm$0.09 &                 &   0.27$\pm$0.08 \\
{[Cl III]  } & 5517 & -0.154 &   0.26$\pm$0.06 &                 &   0.37$\pm$0.09 &   0.25$\pm$0.06 &                 &   0.34$\pm$0.09 \\
{[Cl III]  } & 5537 & -0.157 &   0.14$\pm$0.05 &                 &                 &   0.13$\pm$0.04 &                 &                 \\
{N II      } & 5667 & -0.180 &   0.63$\pm$0.05 &                 &   0.83$\pm$0.09 &   0.58$\pm$0.05 &                 &   0.77$\pm$0.08 \\
{N II      } & 5676 & -0.181 &   0.29$\pm$0.05 &                 &   0.18$\pm$0.15 &   0.27$\pm$0.05 &                 &   0.16$\pm$0.13 \\
{N II      } & 5680 & -0.182 &   1.14$\pm$0.07 &                 &   1.56$\pm$0.14 &   1.05$\pm$0.06 &                 &   1.43$\pm$0.13 \\
{N II      } & 5686 & -0.183 &   0.29$\pm$0.04 &                 &   0.22$\pm$0.08 &   0.26$\pm$0.04 &                 &   0.21$\pm$0.07 \\
{N II      } & 5710 & -0.187 &   0.26$\pm$0.04 &                 &   0.19$\pm$0.06 &   0.24$\pm$0.04 &                 &   0.18$\pm$0.06 \\
{[N II]    } & 5754 & -0.194 &   1.43$\pm$0.07 &                 &   1.23$\pm$0.08 &   1.31$\pm$0.06 &                 &   1.12$\pm$0.07 \\
{He I      } & 5876 & -0.215 &  32.80$\pm$0.30 &  39.10$\pm$2.75 &  33.30$\pm$0.90 &  29.71$\pm$0.27 &  35.41$\pm$2.49 &  30.17$\pm$0.82 \\
{[O I]     } & 6300 & -0.282 &   0.66$\pm$0.09 &                 &   0.87$\pm$0.22 &   0.58$\pm$0.08 &                 &   0.76$\pm$0.20 \\
{[S III]   } & 6312 & -0.283 &   0.35$\pm$0.08 &                 &   0.44$\pm$0.22 &   0.31$\pm$0.07 &                 &   0.38$\pm$0.19 \\
{[O I]     } & 6363 & -0.291 &   0.19$\pm$0.08 &                 &   0.16$\pm$0.16 &   0.17$\pm$0.07 &                 &   0.14$\pm$0.14 \\
{C II      } & 6462 & -0.306 &   0.85$\pm$0.08 &                 &   0.92$\pm$0.20 &   0.74$\pm$0.07 &                 &   0.80$\pm$0.18 \\
{[N II]    } & 6548 & -0.318 &  15.00$\pm$0.70 &  18.80$\pm$3.98 &  15.00$\pm$0.70 &  12.96$\pm$0.60 &  16.24$\pm$3.44 &  12.96$\pm$0.60 \\
{H I       } & 6563 & -0.320 & 421.00$\pm$8.00 & 421.00$\pm$7.48 & 421.00$\pm$16.0 & 363.29$\pm$6.90 & 363.31$\pm$6.46 & 363.29$\pm$13.8 \\
{[N II]    } & 6584 & -0.323 &  47.90$\pm$1.30 &  58.60$\pm$4.12 &  46.40$\pm$1.40 &  41.28$\pm$1.12 &  50.50$\pm$3.55 &  39.98$\pm$1.21 \\
{He I      } & 6678 & -0.336 &   9.72$\pm$0.16 &                 &   9.81$\pm$0.18 &   8.33$\pm$0.14 &                 &   8.40$\pm$0.15 \\
{[S II]    } & 6716 & -0.342 &   5.76$\pm$0.16 &                 &   5.66$\pm$0.13 &   4.92$\pm$0.14 &                 &   4.84$\pm$0.11 \\
{[S II]    } & 6731 & -0.344 &   5.18$\pm$0.17 &                 &   5.00$\pm$0.13 &   4.42$\pm$0.15 &                 &   4.27$\pm$0.11 \\
{He I      } & 7065 & -0.387 &   2.70$\pm$0.24 &                 &   3.11$\pm$0.19 &   2.26$\pm$0.20 &                 &   2.60$\pm$0.16 \\
{[Ar III]  } & 7135 & -0.396 &   7.70$\pm$0.25 &                 &   7.52$\pm$0.20 &   6.42$\pm$0.21 &                 &   6.27$\pm$0.17 \\
\end{tabular}
\end{minipage}
\end{table*}

An initial assessment of the data shows Hf\,2-2 to be a fairly low density
nebula, as indicated by the observed [O~{\sc ii}] and [S~{\sc ii}] doublet
ratios. The extinction towards Hf\,2-2 also seems to be low. We thus
carried out a preliminary plasma disgnostic analysis assuming zero
interstellar reddening towards Hf\,2-2. This yields He$^+$/H$^+$ and
He$^{++}$/H$^+$ abundance ratios of 0.127 and 0.002 from the He~{\sc i}
$\lambda$4471 and He~{\sc ii} $\lambda$4686 lines, respectively. These He
ionic abundances, together with the observed ratio of the Balmer
discontinuity to H\,11 of 0.569~\AA $^{-1}$, yield a Balmer jump electron
temperature of $T_{\rm e}({\rm BJ}) = 890$~K [c.f. Eq. (3) of Liu et al.
(2001)]. The observed [O~{\sc ii}] $\lambda\lambda$3726, 3729 and [S~{\sc
ii}] $\lambda\lambda$6716, 6731 doublet ratios of 1.43$\pm$0.06 and
1.12$\pm$0.03, which are unaffected by reddening, yield electron densities
of ${\rm log}\,N_{\rm e} = 3.1$ and 2.5 (cm$^{-3}$), respectively. In what
follows, $T_{\rm e} = 900$~K and $N_{\rm e} = 1000$~cm$^{-3}$ will be
adopted in order to derive the extinction towards Hf\,2-2 from a
comparison of the observed Balmer decrement with recombination theory
(Storey \& Hummer 1995) and from the ratio of the observed H$\beta$ to
radio free-free fluxes.

The INT spectrum, obtained with a slitwidth of 4\,arcsec, yielded an
H$\alpha$/H$\beta$ ratio 4.21, compared to the predicted value of 3.63 for
$T_{\rm e}$ = 900~K, yielding a logarithmic extinction at H$\beta$ of $c =
0.20$, for a standard Galactic reddening law (Howarth 1983). Although all of
the ESO spectra were obtained under photometric conditions, the blue and red
spectra were observed on different nights and the pointing accuracy of the ESO
1.52-m telescope and the positioning of the Boller \& Chivens spectrograph
long-slit cannot be guaranteed to have repeated exactly on different nights.
From the absolute fluxes of H$\gamma$ and H$\alpha$ recorded on the ESO blue
and red spectra, respectively (H$\beta$ in the ESO red spectra was unreliable
due to vignetting), we derive reddening constants $c = 0.20$ and 0.09 from the
2 and 8\,arcsec slitwidth observations, respectively. Note that in this
particular nebula, the O~{\sc ii} ORLs are so strong that approximately 2 per
cent of the observed flux of H$\gamma$ is due to several O~{\sc ii} lines that
are blended with it.

Compared to the small redddening implied by the observed INT H$\alpha$/H$\beta$
ratio (consistent with the values derived from the H$\alpha$/H$\gamma$ ratios
from the ESO blue and red spectra, relying on the absolute flux calibration),
the fluxes of higher order Balmer lines in the ESO blue spectra, H\,9
$\lambda$3835, H$\epsilon$ $\lambda$3970 and H$\delta$ at $\lambda$4101,
relative to H$\gamma$, yield consistently higher extinction values -- from the
2\,arcsec slit observations, we find c(H$\beta$) = 0.67, 0.67 and 0.72 from the
H\,9/H$\gamma$, H$\epsilon$/H$\gamma$ and H$\delta$/H$\gamma$ ratios,
respectively. Given the short baseline in wavelength, these values are
sensitive to small errors in the measured line ratios -- even in the case of
H\,9/H$\gamma$, a 5 per cent uncertainty in the ratio translates into an error
of 0.2 in c(H$\beta$), the consistently higher reddening derived from these
high order Balmer lines compared to that yielded by the H$\alpha$/H$\beta$
ratio is difficult to explain by observational uncertainies alone. One
possibility is that part of the extinction seen towards Hf\,2-2 is local and
those grains have a steeper extinction curve at shorter wavelength compared to
the standard curve for the diffuse interstellar medium. Kaler (1988) derived a
value of $c({\rm H}\beta) = 0.88\pm 0.06$ from the observed ratios of high
order Balmer lines relative to H$\beta$. Kaler's observations were obtained at
very large airmasses. 
% In spite of the small error quoted, his value is probably insecure given the
% small baseline in wavelength and the very large airmasses under which his
% observations were obtained
Nonetheless, for $T_{\rm e} = 10^4$~K as assumed by Kaler, the higher order
Balmer lines measured in our spectra yield an average $c({\rm H}\beta) = 0.82$,
almost identical to Kaler's value.

The radio free-free continuum flux of Hf\,2-2 at 1.4\,GHz has been measured by
Condon \& Kaplan (1998), who give $S(1.4\,{\rm GHz}) = 4.4\pm 0.4$\,mJy. This,
together with the total H$\beta$ flux of ${\rm log}\,F({\rm H}\beta) = -11.95$
(ergs\,cm$^{-2}$\,s$^{-1}$) derived in the previous subsection, and the He
ionic abundances given above, yields a reddening constant of $c({\rm H}\beta) =
0.50$. Given the uncertainties in the derived total H$\beta$ flux and in the
radio continuum flux, this value for the reddening constant is probably only
accurate to 0.1~dex\footnote{This interpretation of the H$\beta$ to the radio
continuum flux ratio as a reddening diagnostic is actually questionable, given
the extremely peculiar thermal structure of Hf\,2-2, as indicated by the
enormous difference between the electron temperatures drived from the hydrogen
Balmer jump and from the [O~{\sc iii}] forbidden line ratio. In Hf\,2-2, the
Balmer jump temperature is so low that the hydrogen optical continuum emission
($\propto T_{\rm e}^{-1.5}$), optical line emision ($\propto T_{\rm e}^{-0.8}$)
and radio continuum emission ($\propto T_{\rm e}^{-0.5}$) could have
significantly different average emission temperatures.}.

We will adopt $c({\rm H}\beta) = 0.20$ for our analysis. The small
uncertainties in the amount of reddening towards Hf\,2-2 hardly affect our
analysis given that most of the ORL diagnostic lines used have wavelengths
close to H$\beta$. The interstellar extinction maps of Schlegel, Finkbeiner \&
Davis (1998) predict $E({\rm B-V}) = 0.33$ for the direction towards Hf~2-2,
corresponding to $c({\rm H}\beta) = 0.47$, implying that Hf~2-2 lies less than
half-way along the Galactic dust column in this direction.

\section{Properties of the central star} 

\begin{figure} \centering \epsfig{file=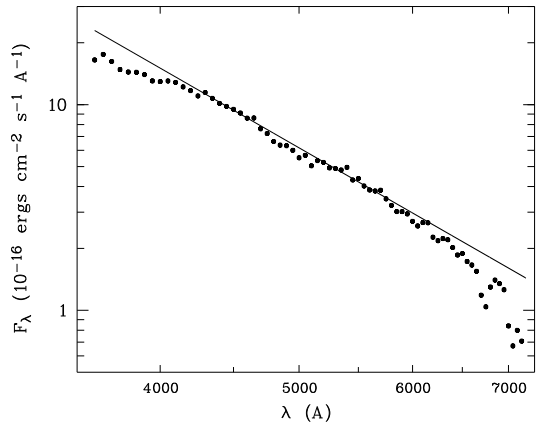, width=7.0cm, bbllx=50pt,
bblly=121pt, bburx=541pt, bbury=513pt, clip=, angle=0}  \caption {Spectrum of
the central star binned in wavelength steps of 50~{\AA}.  The solid line 
shows the energy distribution of a 67,000~K blackbody.} \end{figure}

Hf~2-2 has an unusual central star (CS) -- Lutz et al. (1998) have reported it
as a photometric variable, with a period of 0.398571 days, though no further
information is available. To estimate the magnitude of the CS at the time of
our observations, we made use of the blue and red spectra obtained with
slitwidths of 4 and 8\,arcsec, respectively. The blue spectrum obtained with
an 8\,arcsec slitwidth was not used as it was obtained under very poor seeing
conditions.

In spite of the sub-arcsec seeing conditions under which the two blue and red
spectra were obtained, due to the poor tracking accuracy of the ESO 1.52-m
telescope and its guiding system, the FWHM of the CS at 3930~{\AA} and
5100~{\AA} in the slit direction as determined directly from the spectra were
found to be 2.7 and 3.2 pixels, or 2.2 and 2.6\,arcsec, respectively. Nine CCD
rows centred on the CS, equivalent to a width of 7.3\,arcsec, were co-added to
obtain the integrated spectra of the CS. The blue and red spectra overlap
between 4750--4810~{\AA}, where the continuum level of the red spectrum is
found to be 37 per cent higher than that of the blue spectrum. We therefore
multipled the blue spectrum flux-levels by 1.37 so that the blue and red
spectra joined smoothly. To obtain the fluxes for the CS alone, we need however
to correct for the contribution from the underlying nebular emission to the
extracted CS spectrum.

If the physical conditions (electron temperature, density and the He ionic
abundances relative to H$^+$) in the ionized region where the nebular continuum
emission orginates are known, as well as the reddening towards the nebula, the
contribution of nebular continuum emission to the integrated central star
spectrum can be calculated and then subtracted with the aid of recombination
theory. One uncertainty in this approach is the unknown efficiency of the
H~{\sc i} two-photon emission, which is an important source of nebular
continuum emission shortwards of 4000~{\AA}.  Here we have opted to correct for
the nebular continuum emission to the integrated CS spectrum by using the
nebular spectrum sampled directly from the bright nebular shell. Unfortunately,
a field star, at a distance of 5.11\,arcsec from the CS at PA$=32.6^{\rm o}$,
fell into our 4 and 8\,arcsec slits positioned at PA$=45^{\rm o}$,
contaminating the continuum spectrum of the north-east parts of the nebula. The
field star is about 1.4\,mag fainter than the CS in the [O~{\sc iii}] image
obtained by Schwarz et al.  Thus only the south-west nebular region was used.
Three CCD rows, sampling the nebula 3.8 to 7.0\,arcsec from the CS were
co-added. The resultant blue and red spectra were scaled, using H$\gamma$ and
H$\alpha$, respectively, to the corresponding line fluxes seen in the extracted
CS spectra and then subtracted from the latter.

After correcting for the underlying nebular continuum emission in the
integrated CS spectra, residual emission from some nebular emission lines,
including the prominent C~{\sc ii} and O~{\sc ii} recombination lines, is
present in the corrected CS spectrum, suggesting that the intensities of these
lines relative to H~{\sc i} Balmer lines, are stronger towards the nebular
centre. Similarly, some residual Balmer continuum emission was present
shortwards of the 3646~{\AA} limit, indicating that the central regions of the
nebula may have a lower Balmer jump temperature.  Spatial variations of the
nebular properties will be discussed in detail in the next section.

No spectral features are found that can be attributed to the CS. The CS blue
and red spectra were merged and binned in wavelength steps of 50~{\AA}. The
resulting spectrum is plotted in Fig.\,3. The observed fluxes at 4400 and
5500~{\AA} are 10.4 and $4.4\times
10^{-16}$\,ergs\,cm$^{-2}$\,s$^{-1}$\,{\AA}$^{-1}$. After convolving the
spectral energy distribution (SED) with standard B- and V-band response curves,
we found B = 17.04 and V = 17.37 and a colour index ${\rm B}-{\rm V}=-0.33$
(Kaler 1988 derived B = 17.5 and V = 18.2). The observed color index is
comparable with values found for the hottest {\em unreddened} stars, indicating
that the extinction towards the CS of Hf\,2-2 is negligibly small, $E_{\rm B-V}
\leq 0.1$, or equivalently $c \leq 0.15$, which is consistent with the very low
reddening derived from the H$\alpha$/H$\beta$ ratios in {\S 4}.

Our measured central star flux of $F(\lambda5500) =
4.4\times10^{-16}$\,ergs\,cm$^{-2}$\,s$^{-1}$\,{\AA}$^{-1}$, combined with the
dereddened nebular H$\beta$ flux of
1.78$\times10^{-12}$~ergs~cm$^{-2}$~s$^{-1}$, yields a blackbody H~{\sc i}
Zanstra temperature of 56,000~K, while the above combined with the He~{\sc ii}
$\lambda$4686/H$\beta$ dereddened flux ratio of 0.028 measured with the
8\,arcsec slit implies a blackbody He~{\sc ii} Zanstra temperature of 67,000~K.
If we had instead dereddened the stellar V magnitude using $c({\rm H}\beta) =
0.20$, corresponding to $A_{\rm V} = 0.43$, we would have obtained H~{\sc i}
and He~{\sc ii} Zanstra temperatures of 50,000~K and 64,000~K, respectively.
That the He~{\sc ii} Zanstra temperature is higher than the H~{\sc i} Zanstra
temperature, is consistent with this low-density nebula being optically thin in
the H~{\sc i} Lyman continuum.

In Fig.\,3, we have plotted the SED of an unreddened blackbody of 67,000~K,
normalized to the observed SED of the CS of Hf\,2-2 at 5500~{\AA}. The two SEDs
agree well with each other between 4200--6500~{\AA}. The decline of the
observed SED beyond 6500~{\AA} is clearly caused by the very low S/N at these
long wavelengths, and the sharp drop of fluxes near 7000~{\AA} is due to
absorption by the telluric atmospheric a- and B-band (at 7160 and 6867\,{\AA},
respectively). There is however a significant departure of the observed SED
from that of a 67,000~K blackbody shortwards of 4200~{\AA}, suggesting higher
extinction than at longer wavelengths, which is in line with the higher
reddening derived from the ratios of higher order Balmer lines to H$\gamma$. It
seems to us that the discrepancy between the reddenings derived from the blue
and red wavelength regions is unlikely to be caused by observational
uncertainties and is thus probably real.

\section{Plasma diagnostics and abundance analyses}

\subsection{CEL analysis}

\setcounter{table}{2}
\begin{table}
\begin{minipage}{85mm}
\centering
\caption{Electron temperatures, densities and ionic and elemental abundances 
deduced from collisionally excited lines}
\begin{tabular}{lccc}
Diagnostic & \multicolumn{3}{c}{Slit width}\\
           & 2$\arcsec$ & 4$\arcsec$ & 8$\arcsec$ \\
\noalign{\vskip3pt}
           & \multicolumn{3}{c}{$T_{\rm e}$ (K)}\\
{[O~{\sc iii}]} $(\lambda4959 + \lambda5007)/\lambda4363$ & 8740  & 9060 &  8910 \\
\noalign{\vskip2pt}
{[N~{\sc ii}]} $(\lambda6548 + \lambda6584)/\lambda5754$  & 14460$^a$ &      & 13400$^b$ \\
\noalign{\vskip3pt}
           & \multicolumn{3}{c}{$N_{\rm e}$ (cm$^{-3}$)}\\
{[O~{\sc ii}]} $\lambda3729/\lambda3726$ & 1600  & 1200 &  1600 \\
\noalign{\vskip2pt}
{[S~{\sc ii}]} $\lambda6731/\lambda6716$ & 360   &      & 330 \\
\noalign{\vskip3pt}
           & \multicolumn{3}{c}{$10^6\times$N$^+$/H$^+$}\\
{[N~{\sc ii}]} $\lambda\lambda$6548,6584 & 9.74 & 12.0 & 9.50 \\
           & \multicolumn{3}{c}{N/H = $5.89\times 10^{-5}$}\\
\noalign{\vskip3pt}
           & \multicolumn{3}{c}{$10^5\times$O$^+$/H$^+$}\\
{[O~{\sc ii}]} $\lambda\lambda$3726,3729 & 2.13 & 2.25 & 2.04 \\
           & \multicolumn{3}{c}{$10^4\times$O$^{++}$/H$^+$}\\
{[O~{\sc iii}]} $\lambda\lambda$4959,5007 & 1.05 & 1.09 & 1.08 \\
           & \multicolumn{3}{c}{O/H = $1.28\times 10^{-4}$}\\
\noalign{\vskip3pt}
           & \multicolumn{3}{c}{$10^5\times$Ne$^{++}$/H$^+$}\\
{[Ne~{\sc iii}]} $\lambda\lambda$3868,3967 & 3.60 & 3.60 & 3.42 \\
           & \multicolumn{3}{c}{Ne/H = $4.20\times 10^{-5}$}\\
\noalign{\vskip3pt}
           & \multicolumn{3}{c}{$10^7\times$S$^+$/H$^+$}\\
{[S~{\sc ii}]} $\lambda\lambda$6716,6731 & 3.46 &                      & 3.38 \\
           & \multicolumn{3}{c}{$10^6\times$S$^{++}$/H$^+$}\\
{[S~{\sc iii}]} $\lambda\lambda$6312     & 1.25 &                      & 1.53 \\
           & \multicolumn{3}{c}{S/H = $2.32\times 10^{-6}$}\\
% \noalign{\vskip3pt}
%            & \multicolumn{3}{c}{Cl$^{++}$/H$^+$}\\
% {[Cl~{\sc iii}]} $\lambda\lambda$5517     & $4.43\times 10^{-8}$ &                      & $6.02\times 10^{-8}$ \\
%            & \multicolumn{3}{c}{Cl/H = $6.24\times 10^{-8}$}\\
\noalign{\vskip3pt}
           & \multicolumn{3}{c}{$10^7\times$Ar$^{++}$/H$^+$}\\
{[Ar~{\sc iii}]} $\lambda\lambda$7135 & 7.27 &                      & 7.10 \\
           & \multicolumn{3}{c}{Ar/H = $1.34\times 10^{-6}$}\\
\noalign{\vskip3pt}
\end{tabular}
\begin{description}
\item $^a$ 11,750~K after correcting for the recombination contribution to the 
observed $\lambda$5754 line flux; $^b$ 10,700~K after correcting for the 
recombination contribution to the observed $\lambda$5754 line flux.
\end{description}
\end{minipage}
\end{table}

Table~3 presents electron temperatures derived from the dereddened [O~{\sc
iii}] $(\lambda4959+\lambda5007)/\lambda4363$ and [N~{\sc ii}]
$(\lambda6548+\lambda6584)$/$\lambda5754$ nebular to auroral line ratios
measured with the different slitwidths, using a mean electron density of
1000~cm$^{-3}$, as derived from the [O~{\sc ii}] $\lambda3729/\lambda$3726 and
[S~{\sc ii}] $\lambda6731/\lambda$6716 doublet ratios. We adopt $T_{\rm
e}$([O~{\sc iii}]) = 8820~K from the [O~{\sc iii}] measurements. The [N~{\sc
ii}] ratios measured with the 2 and 8\,arcsec slitwidths formally give $T_{\rm
e}$([N~{\sc ii}]) = 14,500~K and 13,400~K, respectively, much higher than the
$T_{\rm e}$([O~{\sc iii}]) values.  This is due to a significant contribution
by N$^{2+}$ recombination to the $\lambda$5754 line.  Using Eq.\,(1) of Liu et
al.  (2000), together with the N$^{2+}$/H$^+$ ratio obtained from the N~{\sc
ii} ORLs (see below, Table~4) and a temperature of 900~K for the recombining
ions, we predict a recombination contribution of 0.42, on a scale where
H$\beta$=100, to the $\lambda$5754 line, over 30 per cent of its observed
intensity. The $T_{\rm e}$'s derived from the corrected [N~{\sc ii}] ratios are
11,750~K and 10,700~K for spectra taken with the 2\,arcsec and 8\,arcsec
slitwidths, respectively.  Given the significant contamination by recombination
of the $\lambda$5754 line, we will not use $T_{\rm e}$([N~{\sc ii}]) for our
subsequent analysis. Since the O$^{3+}$ ion is not expected to have a
significant abundance in Hf~2-2, we do not expect that the [O~{\sc iii}]
$\lambda$4363 line is affected by a recombination contribution.

Ionic and elemental abundances derived from CELs using $T_{\rm e}$ = 8820~K and
$N_{\rm e} = 1000$~cm$^{-3}$ and ionization correction factors (icf's) taken
from Kingsburgh \& Barlow (1994) are presented in Table~3.

\subsection{ORL analysis}

\setcounter{table}{3}
\begin{table}
\begin{minipage}{85mm}
\centering
\caption{Electron temperatures and ionic and elemental abundances deduced from 
recombination lines and continuum}
\begin{tabular}{lccc}
Diagnostic & \multicolumn{3}{c}{Slit width}\\
           & 2$\arcsec$ & 4$\arcsec$ & 8$\arcsec$ \\
\noalign{\vskip3pt}
           & \multicolumn{3}{c}{$T_{\rm e}$ (K)}\\
BJ/H\,11 & 930 & 875 & 910 \\
% He\,{\sc i} $\lambda$5876/$\lambda$4471 & $940_{-240}^{+240}$  & $< 550$ & $720_{-220}^{+290}$ \\
% He\,{\sc i} $\lambda$6678/$\lambda$4471 & $1170_{-290}^{+320}$ &         & $1000_{-200}^{+210}$ \\
% O\,{\sc ii} $\lambda$4089/$\lambda$4649 & $630_{-380}^{+1280}$ & $630_{-450}^{+2250}$ & $630_{-390}^{+1370}$ \\
He\,{\sc i} $\lambda$5876/$\lambda$4471 & 940  & $< 550$ & 720 \\
He\,{\sc i} $\lambda$6678/$\lambda$4471 & 1170 &         & 1000 \\
O\,{\sc ii} $\lambda$4089/$\lambda$4649 & 630  & 630 & 630 \\
\noalign{\vskip3pt}
           & \multicolumn{3}{c}{He$^+$/H$^+$}\\
He~{\sc i} $\lambda$4471 & 0.102 & 0.102 & 0.101 \\
He~{\sc i} $\lambda$5876 & 0.103 & 0.122 & 0.104 \\
He~{\sc i} $\lambda$6678 & 0.099 &       & 0.100 \\
Adopted                  & 0.102 &       & 0.102 \\
\noalign{\vskip3pt}
           & \multicolumn{3}{c}{He$^{++}$/H$^+$}\\
He~{\sc i} $\lambda$4686 & 0.0018 & 0.0018 & 0.0016 \\
\noalign{\vskip3pt}
           & \multicolumn{3}{c}{He/H = 0.104  }\\
\noalign{\vskip3pt}
           & \multicolumn{3}{c}{C$^{++}$/H$^+$}\\
C~{\sc ii} $\lambda$4267 & $3.68\times 10^{-3}$ & $3.66\times 10^{-3}$ & $3.47\times 10^{-3}$ \\
C~{\sc ii} $\lambda$5342 & $2.98\times 10^{-3}$ &                      & $2.60\times 10^{-3}$ \\
C~{\sc ii} $\lambda$6462 & $3.21\times 10^{-3}$ &                      & $3.47\times 10^{-3}$ \\
Adopted                  & $3.68\times 10^{-3}$ & $3.66\times 10^{-3}$ & $3.47\times 10^{-3}$\\
\noalign{\vskip2pt}
                         & \multicolumn{3}{c}{C/H$ = 4.27\times 10^{-3}$} \\
\noalign{\vskip3pt}
           & \multicolumn{3}{c}{N$^{++}$/H$^+$}\\
N~{\sc ii} $\lambda$5667 & $2.48\times 10^{-3}$ &                      & $3.29\times 10^{-3}$ \\
N~{\sc ii} $\lambda$5676 & $2.60\times 10^{-3}$ &                      & $1.54\times 10^{-3}$ \\
N~{\sc ii} $\lambda$5680 & $2.41\times 10^{-3}$ &                      & $3.28\times 10^{-3}$ \\
N~{\sc ii} $\lambda$5686 & $3.34\times 10^{-3}$ &                      & $2.70\times 10^{-3}$ \\
N~{\sc ii} $\lambda$5710 & $4.64\times 10^{-3}$ &                      & $3.48\times 10^{-3}$ \\
N~{\sc ii} Mult. V\,3& $2.66\times 10^{-3}$ &                      & $3.04\times 10^{-3}$ \\
\noalign{\vskip2pt}
N~{\sc ii} $\lambda$4630 & $2.17\times 10^{-3}$ & $2.78\times 10^{-3}$ & $1.93\times 10^{-3}$ \\
N~{\sc ii} Mult. V\,5& $2.17\times 10^{-3}$ & $2.78\times 10^{-3}$ & $1.93\times 10^{-3}$ \\
\noalign{\vskip2pt}
N~{\sc ii} $\lambda$4035 & $3.19\times 10^{-3}$ & $2.40\times 10^{-3}$ & $2.05\times 10^{-3}$ \\
N~{\sc ii} $\lambda$4041 & $2.78\times 10^{-3}$ & $2.43\times 10^{-3}$ & $2.15\times 10^{-3}$ \\
N~{\sc ii} $\lambda$4044 & $2.11\times 10^{-3}$ & $1.55\times 10^{-3}$ & $2.15\times 10^{-3}$ \\
N~{\sc ii} $\lambda$4176 & $2.89\times 10^{-3}$ & $2.59\times 10^{-3}$ & $2.81\times 10^{-3}$ \\
N~{\sc ii} $\lambda$4237 & $4.61\times 10^{-3}$ & $4.89\times 10^{-3}$ & $4.26\times 10^{-3}$ \\
N~{\sc ii} $\lambda$4242 & $5.66\times 10^{-3}$ & $4.55\times 10^{-3}$ & $4.87\times 10^{-3}$ \\
N~{\sc ii} $\lambda$4433 & $4.71\times 10^{-3}$ & $4.06\times 10^{-3}$ & $4.71\times 10^{-3}$ \\
N~{\sc ii} $\lambda$4530 & $2.04\times 10^{-3}$ & $2.25\times 10^{-3}$ & $1.99\times 10^{-3}$ \\
N~{\sc ii} $\lambda$4552 & $5.40\times 10^{-3}$ & $6.57\times 10^{-3}$ & $5.67\times 10^{-3}$ \\
N~{\sc ii} $\lambda$4678 & $2.64\times 10^{-3}$ & $1.40\times 10^{-3}$ & $2.40\times 10^{-3}$ \\
N~{\sc ii} 3d -- 4f      & $3.33\times 10^{-3}$ & $2.90\times 10^{-3}$ & $2.97\times 10^{-3}$ \\
\noalign{\vskip2pt}
Adopted                  & $2.72\times 10^{-3}$ & $2.84\times 10^{-3}$ & $2.65\times 10^{-3}$ \\
\noalign{\vskip3pt}
           & \multicolumn{3}{c}{N$^{3+}$/H$^+$}\\
N~{\sc iii} $\lambda$4379 & $6.22\times 10^{-5}$ & $8.21\times 10^{-5}$ & $1.02\times 10^{-4}$ \\
\noalign{\vskip2pt}
           & \multicolumn{3}{c}{N/H$ = 3.29\times 10^{-3}$}\\
\end{tabular}
\end{minipage}
\end{table}

\setcounter{table}{3}
\begin{table}
\begin{minipage}{85mm}
\centering
\caption{\it -- continued}
\begin{tabular}{lccc}
Diagnostic & \multicolumn{3}{c}{Slit width}\\
           & 2$\arcsec$ & 4$\arcsec$ & 8$\arcsec$ \\
\noalign{\vskip3pt}
           & \multicolumn{3}{c}{O$^{++}$/H$^+$}\\
O~{\sc ii} $\lambda$4639 & $9.38\times 10^{-3}$ & $9.20\times 10^{-3}$ & $7.68\times 10^{-3}$ \\
O~{\sc ii} $\lambda$4642 & $7.77\times 10^{-3}$ & $8.62\times 10^{-3}$ & $5.90\times 10^{-3}$ \\
O~{\sc ii} $\lambda$4649 & $5.11\times 10^{-3}$ & $4.98\times 10^{-3}$ & $5.11\times 10^{-3}$ \\
O~{\sc ii} $\lambda$4651 & $1.12\times 10^{-2}$ & $1.01\times 10^{-2}$ & $7.86\times 10^{-3}$ \\
O~{\sc ii} $\lambda$4662 & $7.41\times 10^{-3}$ & $8.53\times 10^{-3}$ & $6.71\times 10^{-3}$ \\
O~{\sc ii} $\lambda$4674 & $3.60\times 10^{-3}$ & $3.15\times 10^{-3}$ & $3.60\times 10^{-3}$ \\
O~{\sc ii} $\lambda$4676 & $3.58\times 10^{-3}$ & $3.33\times 10^{-3}$ & $3.67\times 10^{-3}$ \\
O~{\sc ii} Mult. V\,1& $6.63\times 10^{-3}$ & $6.74\times 10^{-3}$ & $5.74\times 10^{-3}$ \\
\noalign{\vskip2pt}
O~{\sc ii} $\lambda$4320 & $3.01\times 10^{-3}$ & $3.23\times 10^{-3}$ &                      \\
O~{\sc ii} $\lambda$4346 & $8.91\times 10^{-3}$ & $8.01\times 10^{-3}$ & $6.99\times 10^{-3}$ \\
O~{\sc ii} $\lambda$4349 & $3.13\times 10^{-3}$ & $4.42\times 10^{-3}$ & $3.37\times 10^{-3}$ \\
O~{\sc ii} $\lambda$4367 & $5.01\times 10^{-3}$ & $7.63\times 10^{-3}$ & $5.96\times 10^{-3}$ \\
O~{\sc ii} Mult. V\,2& $4.55\times 10^{-3}$ & $5.50\times 10^{-3}$ & $4.82\times 10^{-3}$ \\
\noalign{\vskip2pt}
O~{\sc ii} $\lambda$4415 & $7.44\times 10^{-3}$ & $6.06\times 10^{-3}$ & $7.44\times 10^{-3}$ \\
O~{\sc ii} $\lambda$4417 & $1.48\times 10^{-2}$ & $1.08\times 10^{-2}$ & $1.26\times 10^{-2}$ \\
O~{\sc ii} Mult. V\,5& $1.02\times 10^{-2}$ & $7.85\times 10^{-3}$ & $9.39\times 10^{-3}$ \\
\noalign{\vskip2pt}
O~{\sc ii} $\lambda$4072 & $7.43\times 10^{-3}$ & $7.17\times 10^{-3}$ & $6.27\times 10^{-3}$ \\
O~{\sc ii} $\lambda$4076 & $5.06\times 10^{-3}$ & $4.51\times 10^{-3}$ & $4.61\times 10^{-3}$ \\
O~{\sc ii} $\lambda$4079 & $1.07\times 10^{-2}$ & $8.33\times 10^{-3}$ & $9.29\times 10^{-3}$ \\
O~{\sc ii} $\lambda$4085 & $8.14\times 10^{-3}$ & $1.18\times 10^{-2}$ & $1.08\times 10^{-2}$ \\
O~{\sc ii} Mult. V\,10& $6.43\times 10^{-3}$ & $6.16\times 10^{-3}$ & $5.85\times 10^{-3}$ \\
\noalign{\vskip2pt}
O~{\sc ii} $\lambda$4133 & $9.24\times 10^{-3}$ & $5.00\times 10^{-3}$ & $6.82\times 10^{-3}$ \\
O~{\sc ii} $\lambda$4153 & $8.48\times 10^{-3}$ & $9.44\times 10^{-3}$ & $9.65\times 10^{-3}$ \\
O~{\sc ii} $\lambda$4157 & $2.00\times 10^{-2}$ & $2.87\times 10^{-2}$ & $1.00\times 10^{-2}$ \\
O~{\sc ii} $\lambda$4169 & $1.03\times 10^{-2}$ & $6.54\times 10^{-3}$ & $1.22\times 10^{-2}$ \\
O~{\sc ii} Mult. V\,19& $9.84\times 10^{-3}$ & $8.97\times 10^{-3}$ & $9.16\times 10^{-3}$ \\
\noalign{\vskip2pt}
O~{\sc ii} $\lambda$4084 & $7.82\times 10^{-3}$ & $5.56\times 10^{-3}$ & $4.69\times 10^{-3}$ \\
O~{\sc ii} $\lambda$4087 & $8.57\times 10^{-3}$ & $4.28\times 10^{-3}$ & $4.10\times 10^{-3}$ \\
O~{\sc ii} $\lambda$4089 & $5.10\times 10^{-3}$ & $4.95\times 10^{-3}$ & $5.10\times 10^{-3}$ \\
O~{\sc ii} $\lambda$4276 & $6.03\times 10^{-3}$ & $6.27\times 10^{-3}$ & $5.33\times 10^{-3}$ \\
O~{\sc ii} $\lambda$4304 & $1.05\times 10^{-2}$ & $1.07\times 10^{-2}$ & $1.29\times 10^{-2}$ \\
O~{\sc ii} $\lambda$4286 & $8.25\times 10^{-3}$ & $9.04\times 10^{-3}$ &                      \\
O~{\sc ii} $\lambda$4602 & $1.21\times 10^{-2}$ & $1.03\times 10^{-2}$ & $1.00\times 10^{-2}$ \\
O~{\sc ii} $\lambda$4609 & $8.00\times 10^{-3}$ & $9.53\times 10^{-3}$ & $7.88\times 10^{-3}$ \\
O~{\sc ii} 3d -- 4f      & $7.22\times 10^{-3}$ & $6.96\times 10^{-3}$ & $6.63\times 10^{-3}$ \\
\noalign{\vskip2pt}
Adopted                  & $7.48\times 10^{-3}$ & $7.03\times 10^{-3}$ & $6.93\times 10^{-3}$ \\
\noalign{\vskip2pt}
           & \multicolumn{3}{c}{O/H$ = 8.62\times 10^{-3}$}\\
\noalign{\vskip3pt}
           & \multicolumn{3}{c}{Ne$^{++}$/H$^+$}\\
Ne~{\sc ii} $\lambda$4392 & $2.28\times 10^{-3}$ & $2.38\times 10^{-3}$ & $3.21\times 10^{-3}$ \\
Ne~{\sc ii} $\lambda$4220 & $4.29\times 10^{-3}$ & $5.40\times 10^{-3}$ &                      \\
Ne~{\sc ii} 3d -- 4f      & $2.99\times 10^{-3}$ & $3.46\times 10^{-3}$ & $3.21\times 10^{-3}$ \\
\noalign{\vskip2pt}
Adopted                   & $2.28\times 10^{-3}$ & $2.38\times 10^{-3}$ & $3.21\times 10^{-3}$ \\
\noalign{\vskip2pt}
           & \multicolumn{3}{c}{Ne/H$ = 3.28\times 10^{-3}$}\\
\end{tabular}
\end{minipage}
\end{table}

Plasma diagnostics and an abundance analysis using recombination lines and
continuum are presented in Table~4. Electron temperatures have been derived
from the ratio of the Balmer discontinuity at 3646~{\AA} to H\,11 of the H~{\sc
i} spectrum (Liu et al. 2001), from the He~{\sc i} $\lambda$5876/$\lambda$4471
and $\lambda$6678/$\lambda$4471 ratios (Liu 2003; Zhang et al. 2005a,b) and
from the O~{\sc ii} $\lambda$4089/$\lambda$4649 ratio (Liu 2003; Wesson, Liu \&
Barlow 2003). The O~{\sc ii} temperatures were derived by comparing the
observed $\lambda$4089/$\lambda$4649 ratio to theoretical value as a function
of temperature calculated down to a temperature of 288~K (c.f. Fig.\,7 of
Tsamis et al. 2004).

Emissivities of hydrogen recombination lines and continua have only a weak
density dependence at electron densities lower than 10,000\,cm$^{-3}$.  Given
the fairly low density of Hf\,2-2, $\sim 1000$\,cm$^{-3}$ as indicated by
forbidden line density-diagnostics (c.f.  Table~3), precise determinations of
$N_{\rm e}$ using ORLs are difficult. Zhang et al.  (2004) determined electron
temperature and density of H~{\sc i} recombination spectrum by simultaneously
fitting the observed Balmer discontinuity and high-order Balmer lines for a
large sample of Galactic PNe. For Hf\,2-2, they obtained $T_{\rm e}$(BJ) $=
1000\pm 400$~K and $\log N_{\rm e} = 2.6\pm 0.5$\,cm$^{-3}$ using the 2\,arcsec
wide slit spectrum presented here.  The temperature is slightly higher, but
agrees within the uncertainties, with the values presented in Table~4, obtained
using Eq.~(3) of Liu et al. (2001).  Fig.\,4 plots the variation of the He~{\sc
i} $\lambda$5876/$\lambda$4471 and $\lambda$6678/$\lambda$4471 line ratios as a
function of $T_{\rm e}$ for different electron densities. The observed ratios
of Hf\,2-2 indicate that the He~{\sc i} recombination lines must arise from
ionized regions with $N_{\rm e}\la 10^5$\,cm$^{-3}$.
 
\begin{figure} \centering \epsfig{file=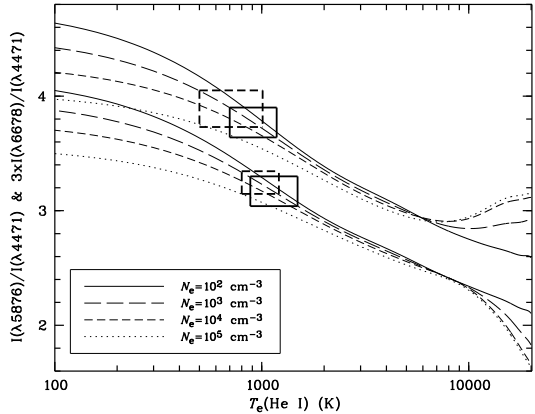, width=7.0cm, bbllx=105pt,
bblly=334pt, bburx=513pt, bbury=655pt, clip=, angle=0}  \caption {Ratios of
He~{\sc i} $\lambda$5876/$\lambda$4471 (the upper set of four curves) and
$3\times \lambda$6678/$\lambda$4471 (the lower set of curves) as a function of
$T_{\rm e}$ for different electron densities. The observed ratios for Hf\,2-2
and the implied temperatures (assuming $N_{\rm e} = 1000$\,cm$^{-3}$) are
marked using dashed-line and solid-line boxes for spectra obtained with the 2
and 8\,arcsec slitwidths, respectively. The sizes of the boxes correspond to
the measurement uncertainties.} \end{figure}

Liu (2003) first used the intensity ratios of the O~{\sc ii} 3d\,$^4$F$_{9/2}$
-- 3p\,$^4$D$^{\rm o}_{7/2}$ $\lambda$4076 (Multiplet V\,10) and
3p\,$^4$D$^{\rm o}_{7/2}$ -- 3s\,$^4$P$_{5/2}$ $\lambda$4649 (Multiplet V\,1)
lines, the strongest recombination lines from the 3d -- 3p and 3p -- 3s
electron configurations, respectively, to the 4f\,G[5]$^{\rm o}_{11/2}$ -
3d\,$^4$F$_{9/2}$ $\lambda$4089 line, the strongest transition from the 4f --
3d configuration, to determine the {\em average}\, electron temperature under
which the lines emitted by Hf\,2-2 arise. This presented the first direct
evidence that they originate from ionized regions with temperatures lower than
1000~K.  On the other hand, Liu (2003) notes that the intensities of other
weaker transitions from the 3d -- 3p and 3p -- 3s configurations, such as the
3d\,$^4$F$_{7/2}$ -- 3p\,$^4$D$^{\rm o}_{5/2}$ $\lambda$4072 and
3p\,$^2$D$^{\rm o}_{5/2}$ -- 3s\,$^2$P$_{3/2}$ $\lambda$4415 lines, appear to
be too strong by $\approx 0.1$ -- 0.2\,dex compared to the $\lambda$4089 line,
yielding higher O~{\sc ii} emission temperatures. Liu (2003) attributed this to
underpopulation of the 2p$^2$\,$^3$P$_2$ level of the recombining O$^{2+}$ ions
relative to the thermal equilibrium value under low nebular densities and
suggested that this effect can be used to measure the {\em average}\, electron
density under which the recombination lines arise. Further evidence pointing to
the underpopulation of the 2p$^2$\,$^3$P$_2$ level of O$^{2+}$ at low densities
is presented by Tsamis et al. (2003a) in their analysis of Galactic and
Magellanic Cloud H~{\sc ii} regions.  

%No calculations of effective recombination coefficients that take into account
%populations of individual fine-structure levels of the recombining and
%recombined ions have been carried out so far. 
Calculations of effective recombination coefficients that take into account the
populations of the individual fine-structure levels of the recombining ion have
now been carried out and preliminary results reported by Bastin \& Storey
(2005). In Fig.\,5 we show the relative intensities of seven of the eight
components of multiplet V\,1 as a function of electron density, from the {\it
ab initio} intermediate coupling calculation of effective recombination
coefficients described by Bastin \& Storey (2005). Also shown are the observed
relative intensities from the 2, 4 and 8\,arcsec slitwidth observations,
positioned at the density that gives the best fit to the theory for the
strongest five components in each case. The best fit electron densities are
4850, 4000 and 13,500\,cm$^{-3}$ for the 2, 4 and 8\,arcsec slitwidth data,
respectively. The electron densities derived from the CELs are significantly
lower, ranging from 330 to 1600\,cm$^{-3}$ and, as can be seen from Fig.\,5, in
strong disagreement with the observed intensity ratios in multiplet V\,1. We
note that the theoretical intensities shown in Fig.\,5 were calculated at
$T_{\rm e} = 10^4$~K (Bastin \& Storey, 2005) while the recombination line
He~{\sc i} and O~{\sc ii} recombination line intensities suggest that an
electron temperature closer to $10^3$~K may be more realistic for the
ORL-emitting region.  The theoretical relative intensities are expected to be
only weakly dependent on temperature in the range $10^3$ -- $10^4$~K so the
conclusions about the density of the ORL-emitting region is expected to remain
valid at the lower temperature. Note that the upper levels of the O~{\sc ii}
$\lambda$4089 and $\lambda$4649 lines have the same O$^{2+}$ 2p$^2$\,$^3$P$_2$
level as parent, so the effect of density on their intensity ratio, and
consequently the O~{\sc ii} ORL temperature deduced from it, is minimal. 

An empirical calibration of the intensity of the 3p\,$^4$D$^{\rm o}_{7/2}$ --
3s\,$^4$P$_{5/2}$ $\lambda$4649 transition relative to the total intensity of
the whole multiplet V\,1 as a function of {\em forbidden line}\, electron
density has also been given by Ruiz et al.  (2003). A more recent calibration,
treating H~{\sc ii} regions and PNe separately, was given by Peimbert \&
Peimbert (2005).  Using their calibration for PNe, we estimate an average
electron density of $N_{\rm e} \la 10^4$\,cm$^{-3}$ for the O~{\sc ii} ORL
emission regions in Hf\,2-2 in broad agreement with the results of the ab
initio calculation.

The ionic abundances presented in Table~4 were derived assuming an electron
temperature $T_{\rm e} = 900$~K and density $N_{\rm e}$ = 1,000\,cm$^{-3}$ for
all ionic species -- hydrogen, helium and heavy elements inclusive. If we
assumed $T_{\rm e} = 630$~K for heavy element ions, as deduced from the O~{\sc
ii} $\lambda$4089/$\lambda$4649 ratio, and $T_{\rm e} = 900$~K for H$^+$, the
resultant ionic abundances of heavy elements are reduced by less than 10 per
cent compared to those tabulated in Table~4. Note that for Ne$^{++}$/H$^+$,
only effective recombination coefficients calculated for a nominal temperature
of 10,000~K are available for the 3d -- 4f transitions. Hence the
Ne$^{++}$/H$^+$ ionic abundances presented in Table~4 should be treated as
quite preliminary.

\begin{figure} \centering \epsfig{file=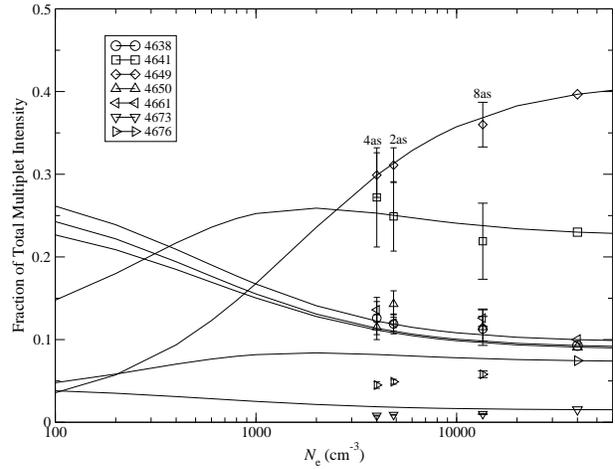, height=8.0cm, bbllx=18pt,
bblly=3pt, bburx=610pt, bbury=769pt, clip=, angle=270}
\caption{Relative intensities of the components of O~{\sc ii} multiplet V1 as a
function of electron density from theory (solid lines) and observations with 2,
4 and 8\,arcsec slits.  The density is varied to obtain the best fit for the
five strongest lines.} \end{figure}

\subsection{Spatial variations}

\begin{table}
\begin{minipage}{85mm}
\centering
\caption{Spatial variations}
\begin{tabular}{lccccc}
Quantity & NE2 & NE1 & Cent & SW1 & SW2 \\
\noalign{\vskip3pt}
  &  &\multicolumn{3}{c}{CEL analysis} & \\ 
$T_{\rm e}$([O~{\sc iii}]) (K) &  &  8750 & 9900 & 8510 &  \\
$N_{\rm e}$([O~{\sc ii}]) (cm$^{-3}$) & 1290 & 1990 & 2550 & 1890 & 860  \\
$N_{\rm e}$([S~{\sc ii}]) (cm$^{-3}$) &  380 &  340 &  840 &  450 & 320  \\
$10^4\times$O$^+$/H$^+$ & 0.155 & 0.210 & 0.240 & 0.210 & 0.213 \\
$10^4\times$O$^{++}$/H$^+$ & 0.851 & 1.25 & 1.17 & 1.21 & 0.732 \\
$10^4\times$O/H & 1.01 & 1.46 & 1.41 & 1.42 & 0.945 \\
N$^+$/O$^+$ & 0.641 & 0.315 & 0.329 & 0.304 & 0.770 \\
\noalign{\vskip3pt}
  &  &\multicolumn{3}{c}{ORL analysis} & \\ 
$T_{\rm e}$(BJ) (K) & 1030 & 870 & 810 & 780 & 1220 \\
$T_{\rm e}$($\lambda5876/\lambda4471$) (K) & 810 & 620 & 910 & 550 & 730 \\
$T_{\rm e}$($\lambda6678/\lambda4471$) (K) & 950 & 900 & 900 & 1010 & 1520 \\
% He$^+$/H$^+$($\lambda$4471) & 0.0976 & 0.0994 & 0.0100 & 0.0991 & 0.0992 \\
% He$^+$/H$^+$($\lambda$5876) & 0.0996 & 0.0104 & 0.0100 & 0.0105 & 0.0102 \\
% He$^+$/H$^+$($\lambda$6678) & 0.0971 & 0.0998 & 0.0101 & 0.0985 & 0.0928 \\
He$^+$/H$^+$ & 0.099 & 0.102 & 0.101 & 0.103 & 0.100 \\
He$^{++}$/H$^+$ &        & 0.002 & 0.004 & 0.002 &        \\
He/H & 0.099 & 0.104 & 0.105 & 0.105 & 0.100 \\
$10^3\times$C$^{++}$/H$^+$ & 2.34 & 4.22 & 5.42 & 3.91 & 1.66 \\ 
$10^3\times$O$^{++}$/H$^+$ & 3.22 & 6.03 & 9.70 & 6.75 & 3.30 \\
C$^{++}$/O$^{++}$ & 0.726 & 0.701 & 0.559 & 0.579 & 0.504 \\
\end{tabular}
\end{minipage}
\end{table}

Given the faintness of the nebula, only limited information regarding the
spatial variations of electron temperature, density and elemental abundances
can be retrieved from the current data set. We divided the nebula into five
regions: 1) NE2 (from $-10.2 < r \le -6.1$\,arcsec, where $r$ is the nebular
radius measured along the slit from the north-east to the south-west, c.f.
Fig.\,2); 2) NE1 ($-6.1 < r \le -2.0$\,arcsec); 3) Centre ($-2.0 < r \le
2.0$\,arcsec); 4) SW1 ($2.0 < r \le 6.1$\,arcsec); and 5) SW2 ($6.1 < r \le
10.2$\,arcsec). Only spectra obtained with the 2\,arcsec slitwidth at the ESO
1.52-m telescope were analyzed. The results are presented in Table~5.  The
first part of the Table presents quantities derived from the CEL analysis, with
consecutive rows giving, respectively, electron temperatures derived from the
[O~{\sc iii}] $(\lambda4959 + \lambda5007)/\lambda4363$ nebular to auroral line
ratio, electron densities deduced respectively from the [O~{\sc ii}]
$\lambda3729/\lambda3726$ and [S~{\sc ii}] $\lambda6731/\lambda6716$ doublet
ratios, ionic abundances O$^+$/H$^+$ and O$^{++}$/H$^+$ derived respectively
from intensities of the [O~{\sc ii}] $\lambda\lambda$3726,3729 lines relative
to H$\gamma$ and of the [O~{\sc iii}] $\lambda\lambda$4959,5007 lines relative
to H$\alpha$, total elemental abundances O/H (sums of O$^+$/H$^+$ and
O$^{++}$/H$^+$), and the ionic abundance ratios N$^+$/O$^+$ with N$^+$/H$^+$
determined from intensities of the [N~{\sc ii}] $\lambda\lambda$6548,6584 lines
relative to H$\alpha$.  All CEL ionic abundances were calculated assuming a
constant electron temperature of 8820~K and an electron density of
1000\,cm$^{-3}$. The [O~{\sc iii}] $\lambda$4363 auroral line was too faint to
be detectable for regions NE2 and SW2. Compared to regions NE1 and SW1, the
central region has an [O~{\sc iii}] temperature about 1000~K higher. From the
outer regions to the centre, we see an increase in electron density by a factor
of two. The total elemental abundance ratio O/H varies by only 40 per cent from
region to region. Given the sensitivity of ionic abundance ratios deduced from
CELs to small errors in temperature, the small variations of O/H are likely to
be caused entirely by measurement uncertainties, in particular those in
temperature determinations. Similarly, the slightly higher N$^+$/O$^+$ ratios
(which, to a good approximation, $\sim {\rm N/O}$) found for the two outermost
regions are probably not real.

In the second part of Table~5, we present plasma diagnostics and ionic and
elemental abundances deduced from ORLs. The consecutive rows give,
respectively, the electron temperatures derived from the ratio of the H~{\sc i}
Balmer jump to H\,11, the temperatures derived respectively from the He~{\sc i}
$\lambda$5876/$\lambda$4471 and $\lambda$6678/$\lambda$4471 ratios, the ionic
abundances He$^+$/H$^+$ and He$^{++}$/H$^+$ determined respectively from the
He~{\sc i} $\lambda\lambda$4471,5876,6678 lines and from the He~{\sc ii}
$\lambda$4686 line, total elemental abundances He/H (sums of He$^+$/H$^+$ and
He$^{++}$/H$^+$), ionic abundance ratios C$^{++}$/H$^+$ and O$^{++}$/H$^+$
derived respectively from the C~{\sc ii} $\lambda$4267 and O~{\sc ii}
$\lambda\lambda$4649,4651 ORLs, and finally the C$^{++}$/O$^{++}$ ionic
abundance ratios.  All ionic abundances were calculated assuming a constant
electron temperature of 900~K and an electron density of 1000\,cm$^{-3}$. The
Balmer jump temperature $T_{\rm e}$(BJ) shows a clear trend of decreasing
towards the nebular centre, whereas no clear trends are found for the He~{\sc
i} temperatures, which are however extremely difficult to measure, given the
very weak dependence of the line ratios on temperature. In dramatic contrast to
the O/H abundance ratios deduced from CELs, both C$^{++}$/H$^+$ and
O$^{++}$/H$^+$ (which, after multiplying by a {\em constant}\, factor of 1.18,
c.f. Table~5, should be very good approximations, within 10 per cent, to C/H
and O/H, respectively) show large enhancements towards the nebular centre, by
as much as a factor of three, with C$^{++}$/O$^{++}$ ($\approx$ C/O) varying by
less than $\sim 40$ per cent.  Previous analyses of NGC\,6153 (Liu et al.
2000), NGC\,6720 (Garnett \& Dinerstein 2001) and NGC\,7009 (Luo \& Liu 2003)
also showed that ORLs strongly peak towards the nebular centre. This thus seems
to be a ubiquitous phenomenon amongst PNe.

\section{Discussion}

\begin{table}
\begin{minipage}{85mm}
\centering
\caption{Comparison of elemetal abundances}
\begin{tabular}{lcccccccc}
Source & He & C & N & O & Ne & S & Ar\\
\noalign{\vskip3pt}
Hf\,2-2 CELs &        &      & 7.77 & 8.11 & 7.62 & 6.36 & 6.13 \\
Hf\,2-2 ORLs & 11.02  & 9.63 & 9.52 & 9.94 & 9.52 &      &      \\
Av. PNe$^a$  & 11.06  & 8.74 & 8.38 & 8.66 & 8.06 & 6.99 & 6.51 \\
Solar$^b$        & 10.90  & 8.39 & 7.83 & 8.69 & 7.87 & 7.19 & 6.55 \\
\end{tabular}
\begin{description}
\item $^a$ Average abundances of Galactic disk and bulge PNe 
(Kingsburgh \& Barlow 1994; Exter, Barlow \& Walton 2004), all based on CEL 
analyses except for helium for which ORLs were used; $^b$ Solar values from 
Lodders (2003)
\end{description}
\end{minipage}
\end{table}

Elemental abundances derived from CELs and from ORLs are compared in Table~6.
Also listed in the Table are average abundances for Galactic disk and bulge PNe
taken from Kingsburgh \& Barlow (1994) and Exter, Barlow \& Walton (2004), and
the latest solar photospheric values compiled by Lodders (2003).

The oxygen abundance derived from the CELs (8.11 on a logarithmic scale where H
= 12.00) is 0.55~dex smaller than the mean value of 8.66 found for Galactic
disk and bulge PNe (Kingsburgh \& Barlow 1994; Exter et al. 2004).  Since it
seems unlikely that Hf~2-2 has an abnormally low oxygen abundance, given the
great strength of its ORLs, we have explored what physical conditions would be
needed to return a more `normal' oxygen abundance from its CELs. We find that
the adoption of $T_{\rm e}$ = 6560~K for both the oxygen CELs and the
recombining H$^+$ in the nebula would yield $\log\,({\rm O/H}) + 12.0 = 8.66$,
while the adoption of $T_{\rm e}$ = 4030~K would bring the oxygen CEL and ORL
abundances into agreement, at $\log\,({\rm O/H}) + 12.0 = 9.94$. If we were to
adopt the $T_{\rm e}$([O~{\sc iii}]) of 8820~K for the oxygen CELs but used
$T_{\rm e}$(BJ) = 900~K, the H~{\sc i} Balmer jump temperature, for H$\beta$,
then we would find $\log\,({\rm O/H}) + 12.0 = 8.88$. However, if H$\beta$ and
the oxygen CELs originate from such physically distinct regions, then an
abundance ratio based on their relative intensities is meaningless.  

There are two ionic species, doubly ionized oxygen and neon, for which
abundances have been determined using both CELs and ORLs. In both cases, the
ORL abundance is about a factor of 70 larger than the CEL value. The abundance
discrepancy factors (adf's) observed in Hf\,2-2 are thus the largest for any
known PN (except, possibly, the H-deficient knots in Abell\,30 where the adf's
reach nearly three orders of magnitude -- c.f. Wesson et al. 2003).  Equally
remarkable is that Hf\,2-2 exhibits the lowest Balmer jump temperature, $\sim
900$~K ever measured for a PN. This is in good agreement with the relation
observed in other PNe, i.e. the adf is positively correlated with the
difference between the [O~{\sc iii}] forbidden line temperature and the Balmer
jump temperature (Liu et al. 2001). The very low Balmer jump temperature of
Hf\,2-2 and its unusually strong heavy element ORLs mimic closely what was
found for the ejected shell around nova DQ Her 1934, for which Williams et al.
(1978) detected a strong broad emission feature at 3644\,{\AA} which they
identified as the Balmer jump formed at a very low temperature not exceeding
about 500~K.

Ever since the detection of a Balmer jump temperature as low as 3560~K for the
Galactic bulge PN M\,1-42, 5660~K lower than the [O~{\sc iii}] forbidden line
temperature measured for the same nebula (Liu et al. 2001), it has become
increasingly clear that PNe, at least those exhibiting large adf's, must
contain another component of previously unknown ionized gas.  This component of
gas is a strong ORL emitter yet is essentially invisible in traditional strong
CELs, as the electron temperature prevailing there is too low to excite any UV
or optical CELs. This is probably caused by the much enhanced cooling by ionic
infrared fine-structure lines that is a consequence of a very high metallicity.
This inference is supported by detailed photoionization modelling
(P\'{e}quignot et al 2003; Tylenda 2003) and by direct measurements of the {\em
average}\, electron temperatures under which various types of line are emitted
(c.f.  Liu 2003, 2005 for recent reviews). The current observation and analysis
add further credibility to this conjecture.  
% new section - pjs
In particular we have provided new observational evidence that the electron
density of the region emitting the ORLs differs significantly from the CEL
emitting region, reinforcing the conclusion that at least two distinct
components are involved in a more complete understanding of the ORL/CEL
spectra.

PNe are found to show a wide range of adf's, from nearly unity, i.e.  agreement
between the CEL and the ORL abundances, up to nearly a factor of 100, as in the
most extreme case reported here.  Two-abundance photoionization modelling of
NGC\,6153 (P\'{e}quignot et al. 2002, 2003), which exhibits an average adf of
approximately 10 (Liu et al. 2000), shows that only a few Jupiter masses of gas
with a metal overabundance of a factor of 100, are needed to account for the
observed intensities of ORLs. In a recent detailed analysis of the He~{\sc i}
recombination spectrum of a large sample of Galactic PNe, Zhang et al. (2005a)
concluded that a typical value of $10^{-4}$ for the filling factor of
hydrogen-deficient material is sufficient to explain the systematic difference
between the electron temperatures determined from H~{\sc i} recombination
spectra and those deduced from the He~{\sc i} ORL ratios.  The latter are found
to be systematically lower than the former, consistent with the expectations of
the two-abundance model but at odds with the traditional paradigms of
temperature fluctuations and/or density inhomogeneities.

The volume emissivity of a recombination line is proportional to the product of
the densities of the emitting ions and of electrons and increases with
decreasing temperature [$\epsilon({\rm X}^{i+},\lambda) \propto N({\rm
X}^{(i+1)+})N_{\rm e}T_{\rm e}^{-\alpha}$, where $N({\rm X}^{(i+1)+})$ is the
density of recombining ions and $\alpha \sim 1$]. The total mass of H-deficient
(metal-rich) material required to produce the observed strengths of ORLs can be
determined if one knows the distance to the nebula as well as the electron
temperature and density of the emitting gas. The distance to Hf\,2-2 is poorly
known. Cahn \& Kaler (1971) and Maciel (1984) gave estimates of 4.0 and
4.5\,kpc, respectively, based on the Shklovsky method (Shklovsky 1956). Given
the peculiar nature of Hf\,2-2, the method is unlikely to be valid. Here we
have simply adopted a nominal value of 4.25\,kpc. If we assume $T_{\rm e} =
630$~K, $N_{\rm e} = 5000$\,cm$^{-3}$ (c.f. \S{6.2}, Table~4 and Fig.\,5), then
from the observed intensity of the O~{\sc ii} $\lambda$4649 line, we find a
total number of O$^{++}$ ions emitting the $\lambda$4649 ORL of
$N_{\lambda4649}({\rm O}^{++}) = 4.8\times 10^{51} (d/4.25\,{\rm kpc})^2$,
where $d$ is the distance to the nebula in kpc. After correcting for the ionic
concentration in O$^+$, the total mass of oxygen equals $7.4\times 10^{-5}
(d/4.25\,{\rm kpc})^2$\,$M_\odot$.  For hydrogen, if we assume $T_{\rm e} =
1000$~K, $N_{\rm e} = 320$\,cm$^{-3}$, as deduced from simultaneously fitting
the observed H~{\sc i} Balmer discontinuity and decrement (Zhang et al. 2004),
we find $N_{\lambda4861}({\rm H}^{+}) = 1.8\times 10^{55} (d/4.25\,{\rm
kpc})^2$, or a total mass of ionized hydrogen $1.4\times 10^{-2} (d/4.25\,{\rm
kpc})^2$\,$M_\odot$. The latter is far less than the nominal total mass of
ionized gas of 0.3\,$M_\odot$ for an optically thin PN, even after taking into
account the contribution from helium.  Similarly, from the [O~{\sc iii}]
$\lambda$5007 forbidden line, assuming $T_{\rm e} = 8920$~K, $N_{\rm e} =
1000$\,cm$^{-3}$ (c.f. \S{6.1}, Table~3), we find $N_{\lambda5007}({\rm
O}^{++}) = 2.8\times 10^{51} (d/4.25\,{\rm kpc})^2$, or a total mass of oxygen
of $4.6\times 10^{-5} (d/4.25\,{\rm kpc})^2$\,$M_\odot$. Thus the amounts of
metal in the cold H-deficient component and in the hot ``normal'' component are
comparable. If we assume that the [O~{\sc iii}] $\lambda$5007 forbidden line
arises entirely from the hot gas, whereas the O~{\sc ii} $\lambda$4649
recombination line originates exclusively from the cold component, then the
total number of oxygen atoms in the nebula is simply the sum of those in the
two components. This leads to an average O/H abundance for the {\em whole}\,
nebula of $5.4\times 10^{-4}$, or, on a logarithmic scale where H = 12.00, O =
8.73, which is almost identical to the solar photospheric value of 8.69. The
O/H abundances of the hot and of the cold gas cannot be determined individually
unless one knows how to separate the contributions from the two components to
the observed total flux of H$\beta$.

Two scenarios have been proposed for the possible origins of the postulated
H-deficient material (Liu 2003; 2005): 1) nucleo-processed material ejected in
a late helium flash as proposed for ``born-again'' PNe such as Abell\,30 and 78
(Iben, Kaler \& Truran 1983); 2) icy material left over from the debris of
planetary system of the progenitor star of the PN. It is also possible that PNe
exhibiting particularly large adf's, including those ``born-again'' PNe, are
related to the phenomenon of novae (Wesson et al. 2003). The last possibility
is particularly attractive in view that Hf\,2-2 is a known close binary system
with an orbital period of only 0.398571\,days (Lutz et al. 1998) and that the
nebula has been suggested to result from a common envelope ejection event
(Soker 1997; Bond 2000).  Further observations, in particular high spectral and
spatial resolution spectroscopic observations using an integral field facility
mounted on a large telescope, may prove critical to discriminate between these
scenarios.

\section{Acknowledgments} We thank Dr. R. Corradi for providing access to the
original [O~{\sc iii}] and H$\alpha$ images of Hf\,2-2 published by Schwarz et
al. (1992). We would also like to thank Dr. R. Rubin for a critical reading of
the manuscript prior to its publication. The work is partly supported by a
joint research grant co-sponsored by the Natural Science Foundation of China
and the UK's Royal Society.

{}

\end{document}